\documentclass[11pt]{article}
\pdfoutput=1
\usepackage{jcappub,natbib}
\input{colordvi.tex}

\def\ga{\mathrel{\raise.3ex\hbox{$>$\kern-.75em\lower1ex\hbox{$\sim$}}}}
\def\la{\mathrel{\raise.3ex\hbox{$<$\kern-.75em\lower1ex\hbox{$\sim$}}}}

\title{Gravitational Wave signatures of inflationary models from \\ Primordial Black Hole Dark Matter}

\author[a]{Juan Garc\'ia-Bellido,}
\author[b]{Marco Peloso,}
\author[b]{Caner Unal}

\affiliation[a]{Instituto de F\'isica Te\'orica UAM-CSIC, Universidad Auton\'oma de Madrid, Cantoblanco,  Madrid, 28049 (Spain)}
\affiliation[b]{School of Physics and Astronomy, and Minnesota Institute for Astrophysics, University of Minnesota, Minneapolis, 55455 (USA)}

\abstract{Primordial Black Holes (PBH) could be the cold dark matter of the universe. They could have arisen from large (order one) curvature fluctuations produced during inflation that reentered the horizon in the radiation era. At reentry, these fluctuations source gravitational waves (GW) via second order anisotropic stresses. These GW, together with those (possibly) sourced during inflation by the same mechanism responsible for the large curvature fluctuations, constitute a primordial stochastic GW background (SGWB) that unavoidably accompanies the PBH formation.  We study how the amplitude and the range of frequencies of this  signal depend on the statistics (Gaussian versus $\chi^2$) of the primordial curvature fluctuations, and on the evolution of the PBH mass function due to accretion and merging. We then compare this signal with the sensitivity of present and future detectors, at PTA and LISA scales. We find that this SGWB will help to probe, or strongly constrain, the early universe mechanism of PBH production. The comparison between the peak mass of the PBH distribution and  the peak frequency of this SGWB will provide important information on the merging and accretion evolution of the PBH mass distribution from their formation to the present era. Different assumptions on the statistics and on the PBH evolution also result in different amounts of CMB $\mu$-distortions. Therefore the above results can be complemented by the detection (or the absence) of $\mu$-distortions with an experiment such as PIXIE. 
}

\begin{document}

\begin{flushright}  IFT--UAM/CSIC-17-056, UMN-TH 3630/17  \end{flushright}

\maketitle
\flushbottom

%%%%%%%%%%%%%%%%%%%%%%%%%%%%%%%%%%%%%
\section{Introduction } 
%%%%%%%%%%%%%%%%%%%%%%%%%%%%%%%%%%%%%

Massive primordial black holes (PBH) could constitute the dominant component of present dark matter, thus resolving one of the remaining mysteries of modern cosmology, see Ref.~\cite{Garcia-Bellido:2017fdg} for a recent review. Several mechanisms have been proposed for their origin and evolution. The most compelling possibility is related to high peaks in the primordial curvature power spectrum, that originated from quantum fluctuations during inflation, which backreact on space-time producing large amplitude curvature fluctuations. These large fluctuations collapse during the radiation era to form black holes with masses of the order of that within the horizon at reentry. If the peak in the curvature power spectrum is broad, subsequent large fluctuations enter close to each other and there is a higher probability that nearby horizons collapse to form PBH, so they are predicted to be clustered today. In this scenario, only a very small  fraction of all causal domains collapse to form black holes. The PBH thus produced constitute only a small fraction of the total energy density during the radiation era, but their relative contribution over radiation grows (nearly) as the scale factor, and comes to dominate at matter-radiation equality. Such PBH could then constitute a considerable fraction of the present matter component. 

The probability of collapse during the radiation era is determined by the amplitude of the curvature perturbations at reentry. A PBH abundance compatible with that of the present dark matter requires perturbations that are significantly greater than the ones measured at the Cosmic Microwave Background (CMB)  scales. While, in typical models of inflaton, CMB modes were generated  approximately 60 $e$-folds before the end of inflation, fluctuations that can lead to present PBH dark matter  were generated  around 40-to-20 $e$-folds before the end of inflation, depending on the precise PBH mass distribution. Various mechanisms for producing PBH have been proposed in the literature, including:  the use of  a scalar field, coupled to the inflaton, with a symmetry breaking potential, triggering a rapid growth of modes during inflation~\cite{GarciaBellido:1996qt,Clesse:2015wea,Clesse:2016vqa}; from the presence of an inflection point in a single-field inflationary potential~\cite{Garcia-Bellido:2017mdw,Ezquiaga:2017fvi,Motohashi:2017kbs,Bezrukov:2017dyv}; from domain walls~\cite{Deng:2016vzb}; from Q-balls~\cite{Cotner:2016cvr,Cotner:2017tir}; from sourcing vector perturbations, amplified by a rolling axion~\cite{Linde:2012bt,Bugaev:2013fya,Erfani:2015rqv,Cheng:2016qzb,Garcia-Bellido:2016dkw,Domcke:2017fix}; or from multiple stages of inflation with an intermediate violation of the slow-roll conditions~\cite{Kannike:2017bxn}.

As a result of these different mechanisms, the amplified density perturbations have different statistical nature (i.e. single-field inflationary models with special features in the potential typically obey Gaussian distributions, but the sourced perturbations may have non-Gaussian, more specifically $\chi^2$, properties.) If the source for the amplified quantum perturbations results from a higher-order interaction, one can have a distribution of the form  $\zeta \propto {\cal G}^n - \langle {\cal G}^n \rangle$, where ${\cal G}$ denotes a Gaussian probability density function. With higher values of the exponent $n$, the PBH production efficiency increases since the probability distribution function becomes more spread, like in Critical Higgs Inflaton,~\cite{Garcia-Bellido:2017mdw,Ezquiaga:2017fvi} (i.e. the region under the tail of the distribution grows).  Since PBH are the result of the very end tail of the probability distribution,\footnote{Note that this is not always true since there are scenarios like those of rapid-waterfall hybrid inflation~\cite{GarciaBellido:1996qt} where the power spectrum is more like a delta function and the tails are suppressed.} smaller amplitude curvature perturbations can produce the same amount of PBH in the case of a non-Gaussian vs. a Gaussian statistics~\cite{PinaAvelino:2005rm}.
 
As noted in Refs.~\cite{Mollerach:2003nq,Ananda:2006af,Baumann:2007zm}, scalar and tensor modes couple to each other at second order in perturbation theory.  Even though this coupling is suppressed by the Planck scale, the enhancement of the scalar perturbations required to produce PBH can induce a significant  amount of gravitational waves (GW). This stochastic GW background (SGWB) is unavoidably present in all models that result in PBH. Its amplitude is very sensitive to both the amplitude of scalar perturbations and their statistics: for an equal abundance of PBH at formation, a smaller SGWB is obtained in the case of non-Gaussian vs. Gaussian primordial curvature modes. The simultaneous detection of the present PBH mass distribution and of the SGWB signal could therefore provide crucial information on the statistics of these modes, and can help discriminating between different models for their production. 

For definiteness, we compare the case of a localized distribution of PBH masses originated from a primordial perturbations that obey $\chi^2$ statistics (as obtained from the specific model \cite{Garcia-Bellido:2016dkw}), vs. the case of a Gaussian distribution. In the former case, we also include the SGWB produced during inflation by the gauge fields that also source the curvature perturbations, and we find that this dominates over the SGWB produced by the curvature modes at reentry.\footnote{In this work we focus on the GW signals generated in PBH models, before or during the PBH formation. In addition to the SGWB considered here, massive PBH (massive enough so not to Hawking-radiate away) can merge during cosmic history and emit GW, generating a unique class of background~\cite{Clesse:2016ajp,Mandic:2016lcn}, which could become an irreducible GWB for LISA~\cite{Bartolo:2016ami}. Furthermore, another class of Stochastic GWB exists due to the non-spherical collapse of PBH which has a peak around similar frequencies but with smaller magnitude, see Refs.~\cite{Garcia-Bellido:2017fdg,Nakama:2016gzw}.} We focus our attention on two mass ranges where, in light of current uncertainties on the PBH limits, a distribution of PBH masses could account for the present dark matter. The most interesting range is that of $M_{\rm PBH} \sim{\cal O} \left( 10 \right) M_\odot$ (where $M_\odot \simeq 2 \times 10^{33} g$ is the solar mass), since collisions of PBH in this mass range  may be responsible for the GW signals observed at LIGO~\cite{Bird:2016dcv,Clesse:2016vqa,Sasaki:2016jop}. In this case, the SGWB produced at reentry is peaked at Pulsar Timing Array (PTA) frequencies,  $f_{\rm peak} \sim {\rm few} \,n$Hz, where the experimental sensitivity is expected to strongly improve with the Square Kilometer Array (SKA) experiment~\cite{Moore:2014lga}. In this work we show that PTA data can significantly probe this mass range of PBH dark matter. 

The relevance of this SGWB at PTA frequencies has also been recently emphasized in  Refs.~\cite{Garcia-Bellido:2016dkw,Orlofsky:2016vbd,Inomata:2016rbd,Nakama:2016gzw}. The two works~\cite{Orlofsky:2016vbd,Inomata:2016rbd} study the SGWB sourced at reentry at PTA scales emerging from different inflationary models, assuming Gaussian statistics of the curvature perturbations. Ref.~\cite{Nakama:2016gzw} computed the peak of this SGWB assuming also Non-Gaussian statistics, and showing how this leads to a decrease of the GW amplitude (at fixed amount of PBH). Contrary to the present work, Ref.~\cite{Nakama:2016gzw} did not compute the scale dependence of this SGWB.  All these works assume a trivial evolution of the PBH from their formation to the present time. In our analysis, we stress the fact that this SGWB probes the PBH mass distribution at the time of its formation. Therefore, as already mentioned in Ref.~\cite{Garcia-Bellido:2016dkw}, the measurement of this signal, together with detailed information on the PBH mass function today, obtained from frequent BH binary (BHB) mergers and close hyperbolic encounters~\cite{Garcia-Bellido:2017qal}, can reveal precious information about the PBH evolution and environment. 

Another experimental handle that can be used to discriminate between the different assumptions (statistics of fluctuations and PBH evolution), is given by the amount of CMB $\mu$-distortions generated in these PBH models~\cite{Clesse:2015wea,Nakama:2016kfq}.  This amount is strongly sensitive on the amplitude and the scale of the bump in the spatial curvature perturbations. We show that the detection (or its absence) of a significant $\mu$-distortion in an experiment such as the Primordial Inflation Explorer (PIXIE)~\cite{PIXIE} or the
Polarized Radiation Imaging and Spectroscopy Mission (PRISM)~\cite{PRISM} can complement what we could learn from the SGWB detection. 

The organization of the paper is as follows: In Section \ref{sec:limits} we discuss the current bounds on PBH. In Section \ref{sec:gaussornongauss} we discuss the significance of statistical properties of perturbations specifically for Gaussian and $\chi^2$ distributions. We introduce two models for these distinct statistics. We devote Section \ref{sec:GW} to the detailed analysis of the contributions to the SGWB from sourced and induced tensor modes produced by these two models. While the production from a Gaussian distribution has been well studied in the literature, the detailed spectrum produced in the non-Gaussian model is an original result of this work. In Section \ref{sec:gaussvsnongauss}, we compare the GW backgrounds of Gaussian and Non-Gaussian models against the expected sensitivity at PTA and LISA scales.  We continue with study of the evolutionary effects in Section \ref{sec:evol}. Finally,  Section \ref{sec:conc} presents our conclusions.

%%%%%%%%%%%%%%%%%%%%%%%%%%%%%%%%%%%%%
\section{Summary of bounds on PBH }
\label{sec:limits}
%%%%%%%%%%%%%%%%%%%%%%%%%%%%%%%%%%%%%

\begin{figure}[tbp]
\centering 
\includegraphics[width=0.47\textwidth,angle=0]{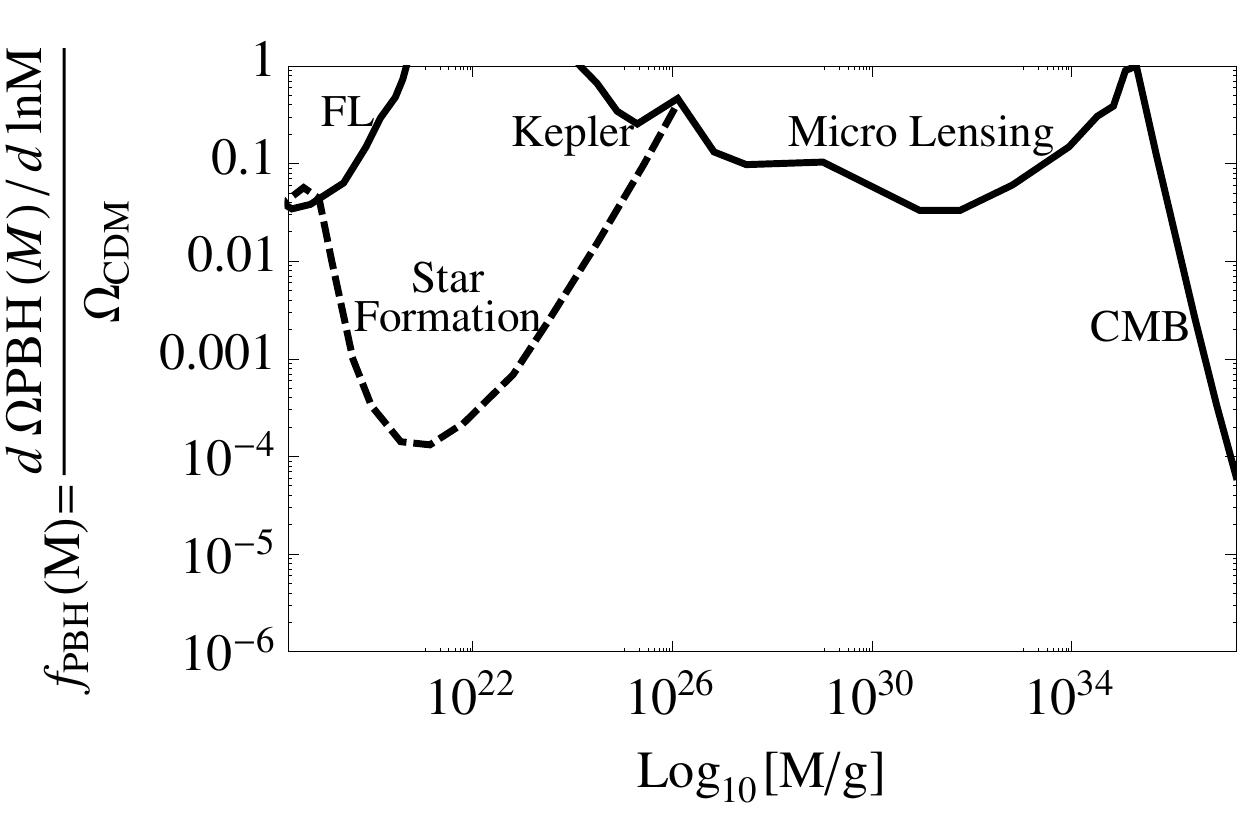}
\includegraphics[width=0.47\textwidth,angle=0]{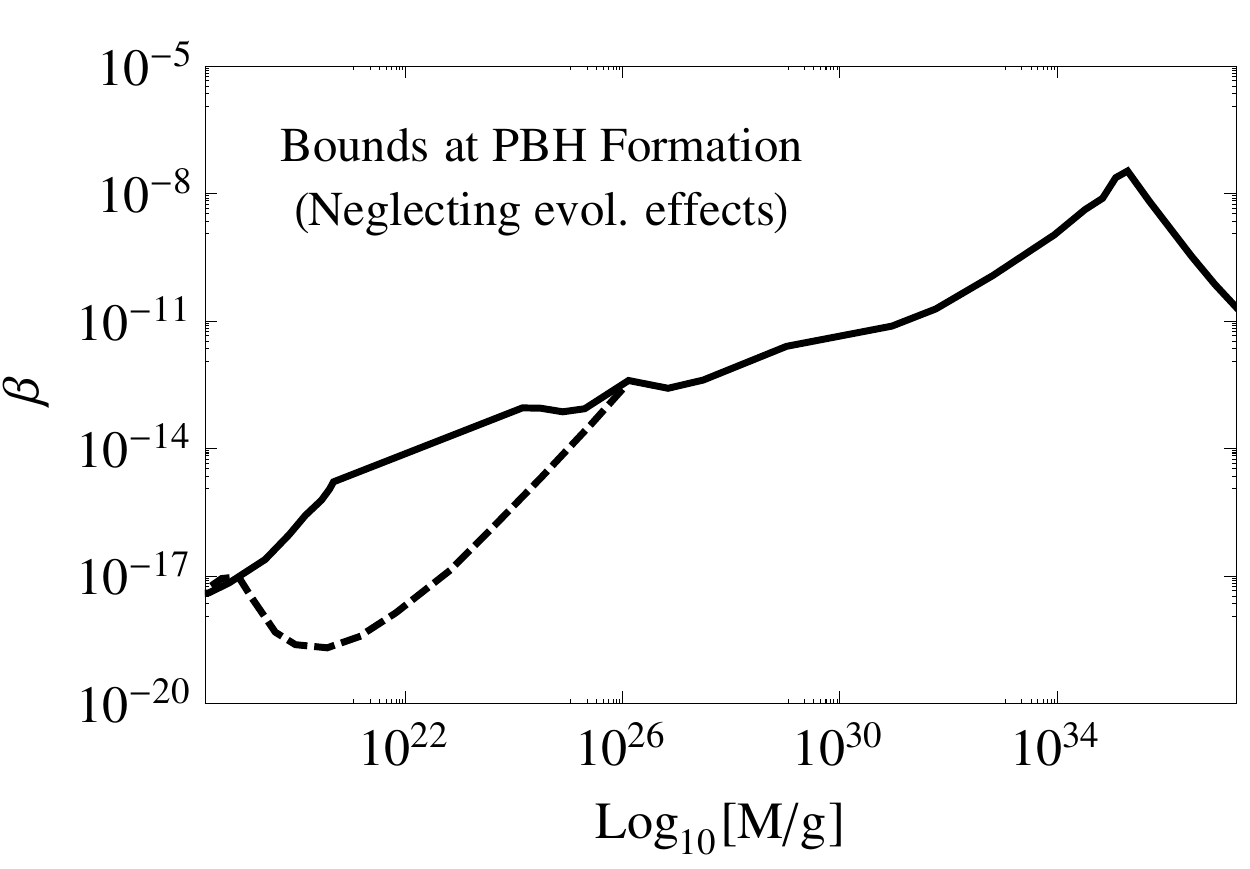}
\hfill
\caption{Limits on the present dark matter fraction in PBH, $f_{\rm PBH}$ (left panel), and on the fraction $\beta$ of regions that collapse to form a black hole (right panel). See the text for details. 
}
\label{fig:limits}
\end{figure}

The left panel of Figure~\ref{fig:limits} is a compilation of  current bounds on the fraction $f_{\rm PBH}$ of dark matter in PBH  as a function of PBH mass (with the $x$-axis ranging from $10^{-15}$ to $10^4~M_\odot$).  Different mass scales have been constrained by different experiments, as we discuss below. The right panel  shows the corresponding limits on the fraction $\beta$ of regions  (of a given size, corresponding to a given black hole mass) that collapse to form a black hole (see Eq.~(\ref{f-A-M}), with  for the relation between $f_{\rm PBH}$ and $\beta$).  We note that evolutionary effects, such as PBH accretion from the surrounding plasma, and merging of PBH, are not included when producing the limits shown in the right panel.\footnote{Namely, we take ${\cal A} = {\cal M}$ = 1 in  Eq.~(\ref{f-A-M}); we discuss evolutionary effects in Section \ref{sec:evol}.}  

Let us review the various constraints, from smaller to larger mass. The femto-lensing (``FL'') line at the lowest masses shown in the figure is due to lack of femto-lensing detection of Gamma-ray bursts by Fermi \cite{Barnacka:2012bm}. The ``Star Formation'' limits are obtained from the capture of PBH dark matter by a star during its formation~\cite{Capela:2014ita}. As discussed in \cite{Kawasaki:2016pql}, there are large uncertainties on these limits, and for this reason we do not include them in our discussion below, where we assume that this mass window can be compatible with PBH being a significant fraction, or the totality, of the dark matter. The ``Kepler'' line is due to the non-observation of microlensing events at Kepler \cite{Griest:2013aaa}.  The ``Micro Lensing'' bounds in the $10^{26} \la M({\rm g}) \la 10^{35}$ range come from observations of microlensing events by the MACHO and EROS Collaborations~\cite{Tisserand:2006zx,Alcock:1998fx}. Both experiments lasted for about six years, and, as a consequence, they cannot constrain higher mass objects.  The mass range $10^{35} \la M({\rm g}) \la 10^{37}$ is mostly constrained by the existence of a stellar cluster near the center of ultra-faint dwarf galaxy (``UFD'') Eridanus II~\cite{Brandt:2016aco}, which has been shown in~\cite{Li:2016utv} to be weakened if there is an intermediate mass black hole of a few thousand solar masses at its center, as the clustered PBH scenario predicts~\cite{Garcia-Bellido:2017fdg}. Therefore, we do not include this bound in our analysis.\footnote{The non-observation of wide binary disruption results in a weaker limit in this mass range~\cite{Quinn:2009zg}.} Finally,  the mass range $M({\rm g}) \ga 10^{35}$, tagged as ``CMB", is constrained by the lack of spectral distortions in the CMB spectrum resulting from the radiation emitted due to accretion on PBH~\cite{Ricotti:2007au,Ali-Haimoud:2016mbv}.\footnote{Ref.~\cite{Ali-Haimoud:2016mbv} improves the analysis first conducted in Ref. \cite{Ricotti:2007au}. The analysis of \cite{Ali-Haimoud:2016mbv} obtains two different limits, depending on different assumption on the effects of the radiation emitted by the PBH on the surrounding gas; we assume the less stringent between these two limits.}

%%%%%%%%%%%%%%%%%%%%%%%%%%%%%%%%%%%%%
\section{Gaussian vs. Non-Gaussian Primordial Overdensities  }
\label{sec:gaussornongauss}
%%%%%%%%%%%%%%%%%%%%%%%%%%%%%%%%%%%%%

The efficiency of PBH formation is strongly dependent on the statistical properties of the primordial overdensitites. Therefore, the PBH bounds given in Figure~\ref{fig:limits} turn into very different upper bounds for the primordial scalar power spectrum, depending on the assumed distributions of the perturbations at those scales. For instance, the fraction $\beta$ of causal regions collapsing onto primordial black holes is related to the power $P_\zeta$ of primordial scalar perturbations by~\cite{Lyth:2012yp,Byrnes:2012yx} 
\begin{equation}
\beta \left( N \right) = \left\{ \begin{array}{l} 
{\rm Erfc} \left(  \frac{\zeta_c}{\sqrt{2 P_{\zeta}(N)}} \right)  \;,\;\;\; {\rm Gaussian \; statistics} \;, \\ 
{\rm Erfc} \left( \sqrt{\frac{1}{2}+\frac{\zeta_c}{\sqrt{2P_{\zeta}(N)} } } \right)  \;,\;\;\; \chi^2 \; {\rm  \; statistics} \; , 
\end{array} \right. 
\label{gauss-chi2}
\end{equation} 
where $\zeta_c$ is the threshold for collapse and  ${\rm Erfc } \left( x \right) \equiv 1 - {\rm Erf } \left( x \right)$ is the complementary error function (see for instance Appendix A.2 of \cite{Garcia-Bellido:2016dkw} for a detailed discussion of the above relations). In these relations, $N$ denotes the number of $e$-folds before the end of inflation at which the density mode that eventually collapses to form a PBH left the horizon during inflation. It is related to the PBH mass through Eq.~(\ref{M-N}).

\begin{figure}[tbp]
\centering 
\includegraphics[width=0.55\textwidth,angle=0]{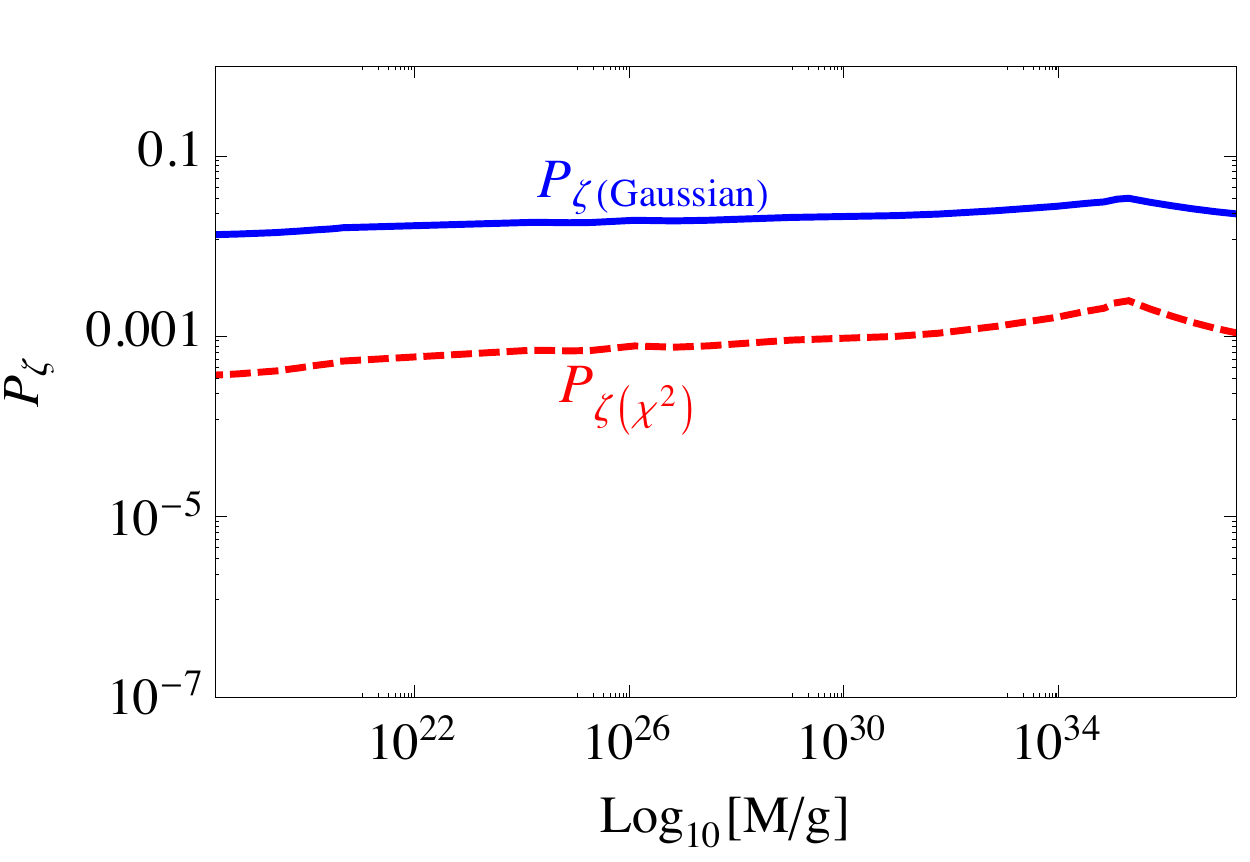}
\hfill
\caption{PBH limits on the power spectrum of the primordial scalar perturbations, assuming that the perturbations obey a Gaussian (solid line) vs. a $\chi^2$ (dashed line) statistics. The limits are obtained from the right panel of Figure \ref{fig:limits}, using Eqs.~(\ref{gauss-chi2}). We note that the power spectrum is much more constrained in the case of a $\chi^2$ statistics, as the perturbations in the tail of the distribution lead to a greater amount of PBH with respect to the Gaussian case. 
}
\label{fig:PSZgauschilimits}
\end{figure}

A given value of $\beta$ corresponds to a very different power in the two cases considered in (\ref{gauss-chi2}). The term $1/2$ in the argument of the second complementary error function can be disregarded for $\zeta_c^2\gg P_\zeta$, which is always satisfied to very good approximation, leading to 
\begin{equation}
P_{ \zeta ( \chi^2)} \simeq \frac{2}{\zeta_c^2} \,  P_{ \zeta \,{\rm (G)}}^2 \,. 
\label{Pchi2-PG}
\end{equation} 
This equation relates the values of the power in the two cases that results in the same value of $\beta$.  Using Eqs.~(\ref{gauss-chi2}) we can translate the bounds given in Figure  \ref{fig:limits} into bounds on $P_\zeta$. The resulting limits are shown in Figure \ref{fig:PSZgauschilimits}. The two lines satisfy the relation (\ref{Pchi2-PG}) with great accuracy.

Different inflationary mechanisms considered in the literature leading to PBH are characterized by different statistics of the scalar perturbations. For instance, the perturbations are Gaussian in the mechanism of Ref.~\cite{Garcia-Bellido:2017mdw}, where a suitable inflaton potential provides an enhanced scalar spectrum at some specific scale in a model of single-field inflation. In the case of hybrid inflation models of PBH production~\cite{GarciaBellido:1996qt,Clesse:2015wea}, the statistics of the peak fluctuations deviates from Gaussian due to quantum diffusion. On the other hand, a $\chi^2$ statistics is obtained in the mechanism of scalar field perturbations arising from the coupling between a rolling axion (different from the inflaton) and a vector field during inflation \cite{Namba:2015gja,Garcia-Bellido:2016dkw}.  The motion of the axion induces a gauge field amplification, that in turn sources scalar primordial perturbations and primordial gravitational waves. The axion is assumed to roll only for a finite amount of $e$-folds $\Delta N$ during inflation, resulting in amplified scalar perturbations only at the scales that left the horizon during this period. This provides a localized ``bump" in the primordial perturbations, with a width related to the model parameter $\Delta N$. These perturbations are not statistically correlated with the vacuum ones (those produced by the expansion of the universe in the absence of this axion-gauge field coupling), resulting in a primordial scalar power spectrum of the form \cite{Namba:2015gja} 
\begin{equation}
P_\zeta= P_{\zeta}^{\rm vac} \left( k_{\rm CMB} \right) \, \left( \frac{k}{k_{\rm CMB}} \right)^{n_s-1} + P_{s,{\rm peak}} \times {\rm exp}  \left[   - \frac{  \ln^2 \left(k / k_{s,{\rm peak}} \right) }{2 \sigma_s^2} \right] \,. 
\label{eq:peakedPS}
\end{equation}
In this expression, the first term is the power spectrum of the vacuum perturbations, for which we assume a standard power-law scaling, with amplitude and tilt given by CMB observations. The second term is the sourced signal resulting from the axion-gauge field coupling. It is characterized by three parameters: the position of the peak in (comoving) momentum space, the height of the signal at the peak, and the width. The sourced scalar perturbations are assumed to be dominant around $k_{s,{\rm peak}}$, but negligible away from it (in particular, they are negligible at CMB scales). They are responsible for the formation of PBH, since the vacuum signal is too small. 

The gauge field also sources a ``bump" in the tensor perturbations, resulting in a GW power spectrum 
\begin{equation}
P_{\rm GW}= P_{\rm GW}^{\rm vac} \left( k_{\rm CMB} \right) \, \left( \frac{k}{k_{\rm CMB}} \right)^{n_t} + P_{t,{\rm peak}} \times {\rm exp}  \left[   - \frac{  \ln^2 \left(k / k_{t,{\rm peak}} \right) }{2 \sigma_t^2} \right] \,. 
\label{eq:peakedGW}
\end{equation}
This expression is analogous to Eq.~(\ref{eq:peakedPS}), and the position and width of the scalar and tensor bump are comparable to each other.  In Appendix \ref{sec:model} we summarize the model of Ref.~\cite{Namba:2015gja} and the precise relations between the model parameters and the scalar and tensor power spectra. 

We denote this model as the ``rolling-axion bump model''. The two key points for the phenomenology of this model are that (i) the scalar perturbations obtained from this mechanism obey a $\chi^2$ statistics; (ii) the bump in the scalar spectrum is accompanied by a correlated bump in the tensor spectrum. To assess the relevance of these two features we compare results obtained for this model with results for a model in which:  
\begin{enumerate}

\item scalar perturbations have a power spectrum still given by Eq.~(\ref{eq:peakedPS}), but obey Gaussian statistics; 

\item there is no corresponding bump in the tensor spectrum. Namely the tensor perturbations obey Eq.~(\ref{eq:peakedGW}) with $P_{t,{\rm peak}} = 0$. 

\end{enumerate} 
For brevity we denote this model as the ``Gaussian bump model''.

%%%%%%%%%%%%%%%%%%%%%%%%%%%%%%%%%%%%%
\section{Primordial vs. Induced Gravitational Waves  }
\label{sec:GW}
%%%%%%%%%%%%%%%%%%%%%%%%%%%%%%%%%%%%%

We identify three distinct populations of GW associated with PBH.\footnote{In addition to the signals considered here, there is also the stochastic background from the non-spherical collapse of PBH~\cite{Garcia-Bellido:2017fdg}. This background can be estimated as $\Omega_{\rm nsc, \, 0} = {\cal E} \cdot  \beta \cdot \Omega_{\rm rad,0} $, where  ${\cal E}$ indicates the efficiency of converting the horizon mass during formation of PBH to GW and $\beta$ is the fraction of causal domains that collapse into a PBH. Using the bound $\beta \la 2\times10^{-8}$, from Figure \ref{fig:limits}, we can estimate $\Omega_{\rm nsc, \, 0}\, h^2 \la 10^{-12}  \cdot  {\cal E}$, which is much smaller than the signals studied here, and thus is ignored.}

In order of their formation, they are: 
\begin{enumerate}

\item The GW produced during inflation by the same mechanism that produces the enhanced scalar perturbations that later become PBH at reentry. We refer to this population as the ``primordial GW'', and we denote it as $h_p$.\footnote{These are {\em not} the vacuum tensor fluctuations produced during quasi-de-Sitter inflation, which are negligible on these scales.} 

\item The GW sourced by the enhanced scalar perturbations. This gravitational production is maximized when the scalar modes re-enter the horizon during the radiation dominated era.  We refer to this population as the ``induced GW'', and we denote it as $h_i$. 

\item The GW produced by the merging of PBH binaries, since formation until today~\cite{Clesse:2016ajp,Mandic:2016lcn}. 

\end{enumerate}

In this work we study the first two populations, in the context of the Gaussian bump model and of the rolling axion bump model introduced in the previous section. 

\begin{figure}[tbp]
\centering 
\includegraphics[width=0.5\textwidth,angle=0]{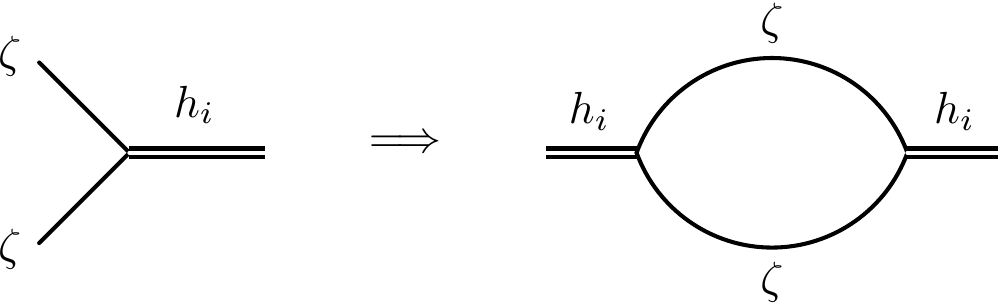}
\hfill
\caption{Diagrammatic expression for the GW induced by scalar perturbations in the Gaussian bump model. 
}
\label{fig:hihi-G}
\end{figure}

The Gaussian bump model assumes that no significant primordial GW are produced. The induced GW are produced by the scalar curvature modes through standard nonlinear gravitational interactions, through a process diagrammatically shown in Figure \ref{fig:hihi-G}. The gravi\-tational interaction is schematically of the type $h \zeta^2$, where $h$ is a tensor mode of the metric (the GW) and $\zeta$ is the scalar curvature (in this schematic discussion we do not indicate the tensorial indices, nor the spatial derivatives acting on $\zeta$, which characterize the interaction).  The tensor mode sourced by this interaction obeys a differential equation that can be solved through a Green function, $G \left( \eta ,\, \eta' \right)$, schematically described as 
\begin{equation}
h_i \left( \eta \right) = \int^\eta d \eta' \ G \left( \eta ,\, \eta' \right) \zeta \left( \eta' \right)  \zeta \left( \eta' \right) \,, 
\label{hzz}
\end{equation}
where $\eta$ is (conformal) time, and where the right hand side contains also a convolution in momenta. This leads to a contribution to the GW power spectrum, schematically as 
\begin{equation} 
\left\langle h_i \left( \eta \right) h_i \left( \eta \right) \right\rangle  = \int^\eta d \eta'  \int^\eta d \eta''  \ G \left( \eta ,\, \eta' \right) 
 G \left( \eta ,\, \eta'' \right) \, \left\langle \zeta \left( \eta' \right)  \zeta \left( \eta'' \right)  \right\rangle  \, 
\left\langle \zeta \left( \eta' \right)  \zeta \left( \eta'' \right) \right\rangle \,. 
\label{hh-zzzz}
\end{equation} 
The two expressions (\ref{hzz}) and (\ref{hh-zzzz}) are diagrammatically shown in Figure \ref{fig:hihi-G}. 

Adding up the two GW polarizations (the induced GW is not polarized, since it is sourced by the scalar $\zeta$), the  total explicit expression corresponding to (\ref{hh-zzzz}) is  \cite{Ananda:2006af}
\begin{eqnarray}
P_{h_i} (\eta, k) &=& \frac{32}{81} \frac{k}{\eta^2}  \int_0^{\eta} d \eta' \int_0^{\eta} d \eta'' \int_0^{\infty} d p \, \int_{-1}^{1} \, dz   \; \frac{p^3   \left ( 1-z^2  \right)^2}{\vert {\bf k - p }\vert^3} \,\, {\cal P}_{\zeta} ( p ) \,\, {\cal P}_{\zeta}( \vert {\bf k-p} \vert ) \,\nonumber\\ \nonumber\\
&& \times \;\; \eta' \; \eta'' \;  \sin(k\eta - k\eta') \, \, \sin(k\eta - k\eta'') \, F_T ( p \, \eta', \vert {\bf k-p} \vert\ \eta' )  \,  F_T ( p \, \eta'', \vert {\bf k-p} \vert\ \eta'' ) \;, \nonumber\\ 
\label{indgwG}
\end{eqnarray}
where ${\bf p}$ is the loop momentum,  z is the cosine of the angle between $ \bf k $ and $ \bf p$, and where
\begin{eqnarray}
F_T (u,v) = 2 T(u) T(v) + \tilde T(u) \tilde T(v) \,, 
\label{F-T} 
\end{eqnarray} 
with 
\begin{eqnarray} 
T (u) =  \frac{9}{u^2} \left[        \frac{\sin(u/ {\sqrt 3}) }{u/ {\sqrt 3}} -\cos (u/ {\sqrt 3})     \right] \,,&&\;\; 
 {\tilde T}(u)=  \frac{3}{u^2} \left[   \frac{(u^2-6) \sin(u/ {\sqrt 3})}{u/ {\sqrt 3}} + 6 \cos (u/ {\sqrt 3})     \right]  \,. \nonumber\\ 
\label{FT-T-Ttil}
\end{eqnarray}

\begin{figure}[tbp]
\centering 
\includegraphics[width=0.5\textwidth,angle=0]{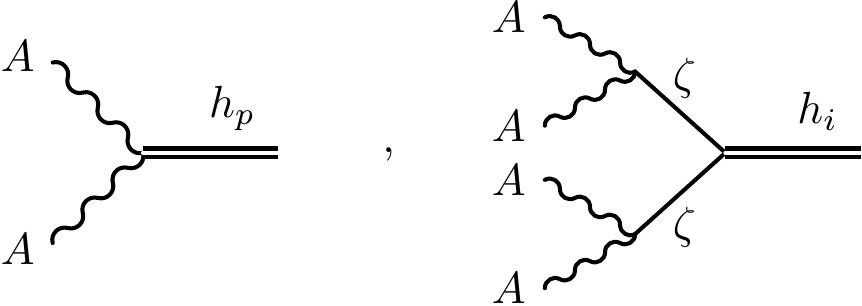}
\hfill
\caption{Primordial and induced GW in the rolling axion bump model. 
}
\label{fig:hphi}
\end{figure}

\begin{figure}[tbp]
\centering 
\includegraphics[width=\textwidth,angle=0]{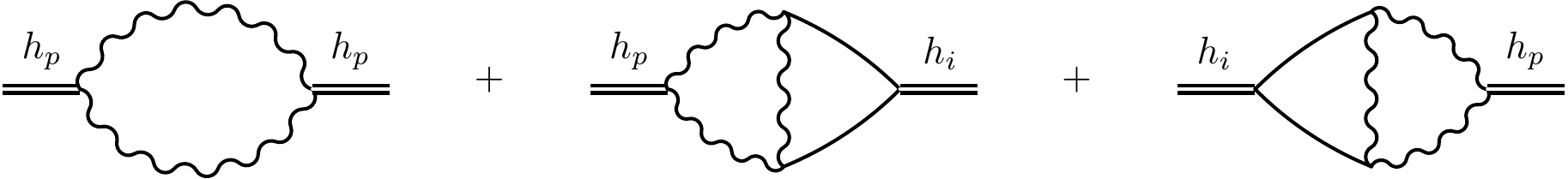}
\hfill
\caption{One and two-loop contributions to the GW signal in the rolling axion bump model. These diagrams give the amplitude of the primordial GW, and of the cross-correlation with the induced GW. Intermediate solid (resp. wiggly) lines represent scalar (resp. gauge field) perturbations. 
}
\label{fig:hphp-hphi}
\end{figure}

Let us now turn our attention to the rolling axion bump model. In this case, both primordial and induced GW are present. 
Figure~\ref{fig:hphi}  shows how the GW  are produced from the vector field $A$ amplified by the rolling axion. The primordial GW are produced by the vector fields during inflation. The autocorrelation $\left\langle h_p h_p \right\rangle$ is of the form (\ref{eq:peakedGW}). This correlator was computed in \cite{Namba:2015gja,Garcia-Bellido:2016dkw}, and it is given by the first diagram of Figure  \ref{fig:hphp-hphi}.

\begin{figure}[tbp]
\centering 
\includegraphics[width=\textwidth,angle=0]{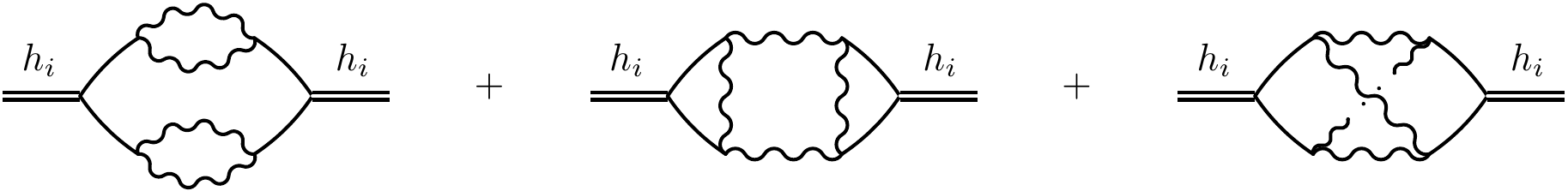}
\hfill
\caption{Auto-correlation of the induced GW signal in the rolling axion bump model. 
Intermediate solid (resp. wiggly) lines represent scalar (resp. gauge field) perturbations. 
}
\label{fig:hihi}
\end{figure}

The induced GWB is produced during the radiation dominated era (mostly at horizon re-entry) by the scalar perturbations that were sourced by the vector fields during inflation.  The induced GW signal in this model was never computed, and it is one of the original results of the present work. Due to the fact that both $h_p$ and $h_i$ originate  from the vector field perturbations,  the total power spectrum $\langle \left( h_p + h_i \right)^2\rangle$ contains also a mixed-term contribution, given by the second and third diagram of Figure \ref{fig:hihi}.

The presence of $h_p$ therefore provides additional contributions to the GW power, that are typically disregarded in works of GW from PBH. Disregarding this signal may not always be a proper assumption, since the production of PBH required a mechanism that enhances the scalar perturbations during inflation, and this mechanism can in principle enhance also the primordial GW. The relevance of $h_p$ over $h_i$ is particularly important in the case in which the scalar perturbations obey Non-Gaussian statistics, as we will show below. The reason for this is that PBH bounds constrain the scalar power much more in the case of Non-Gaussian vs. Gaussian statistics (see Figure \ref{fig:PSZgauschilimits}). This then limits the amount of induced GW which are sourced by these scalar modes. In fact, we will see that $h_p$ dominates over $h_i$ in the rolling axion bump model.

Even if we ignore $h_p$, the study that we perform here constitutes, to the best of our knowledge, the first attempt to fully compute the $\langle h_i h_i \rangle$ auto-correlation in a Non-Gaussian model, where the source of the enhanced scalar perturbations is completely specified.  
In the previous literature, when studying the induced GW in the context of PBH formation, the scalar perturbations are typically assumed to be Gaussian,  so that the source term $\langle \zeta^4 \rangle$ in $\langle h_i h_i \rangle = \int d \eta' d \eta'' G^2 \left\langle \zeta^4 \right\rangle$ can be written as the product of two point functions $P_\zeta^2$, see Eq.~(\ref{indgwG}). In the present context, this Gaussian contribution corresponds to just the first diagram of Figure~\ref{fig:hihi}.  The other two diagrams only emerge when a concrete model is considered, and analogous additional diagrams could be present also for different concrete mechanisms, where {\it e.g.} more fields are involved. 

In general, the 4-point correlator $\left\langle \zeta^4 \right\rangle$ cannot be expressed completely in terms of products of 
2-point correlators  $\left\langle \zeta^2 \right\rangle$, and the expression (\ref{indgwG}) must be replaced by 
\footnote{A prime on a correlator denotes the correlator divided by the corresponding Dirac $\delta$-function.}  
\begin{eqnarray}
P_{h_i} (\eta, k) &\equiv& \frac{k^3}{2 \pi^2} \, \sum_{\lambda=\pm} \left \langle h_{i,\lambda} ( {\bf k}) \,  h_{i,\lambda} ( - {\bf k}) \right \rangle'  \nonumber\\ 
&=&  2 \times \frac{64}{81} \frac{1}{ 2\pi^2} \frac{k}{\eta^2} \int \frac{d^3 p \, d^3 q}{(2\pi)^3} \; p^2 \; \frac{\sin^2 \theta_{\rm kp}}{2} \; q^2 \; \frac{\sin^2 \theta_{\rm k'q}}{2}       \int_0^{\eta} d \eta' \int_0^{\eta} d \eta''   \; \eta' \; \eta'' \;  \sin(k\eta - k\eta') \,     \nonumber\\  \nonumber\\
&& \times   \sin(k' \eta - k' \eta'') \,  \, F_T ( p \, \eta', \vert {\bf k-p} \vert\ \eta' )  \, \, F_T ( q \, \eta'', \vert {\bf k'-q} \vert\ \eta'' )  \;\;  \left  \langle {\hat \zeta}_{\bf p} \,\, {\hat \zeta}_{\bf k-p}\,\,\,\, {\hat \zeta}_{\bf q} \,\, {\hat \zeta}_{\bf k' - q} \right \rangle' \;, \nonumber\\ 
\label{indgwNG}
\end{eqnarray}
where  $F_T$, $T$ and $\tilde T$ are given in Eq.~(\ref{FT-T-Ttil}) and
$\cos\theta_{\rm  kp}={\hat k}\cdot{\hat p}$. Evaluating the $\left\langle \zeta^4 \right\rangle$ correlator in the rolling axion bump model gives rise to the three diagrams shown in Figure \ref{fig:hihi}. The three diagrams are evaluated in Appendix  \ref{appsec:induced}. We denote the first diagram as ``Reducible'', since in this case the  $\left\langle \zeta^4 \right\rangle$ correlator can be reduced to the product of two scalar power spectra. Using the analytic result (\ref{eq:peakedPS}) for the scalar power spectra, this diagram can be evaluated through a one-loop computation. The other two diagrams, which we denote, respectively, as ``Planar" and ``Non-Planar'', must instead be evaluated through a $3$-loop computation. We evaluate them under the approximation of ``zero-width'' gauge modes, namely $\vec{A} \left( \vec{p} \right) \propto \delta \left( \vert \vec{p} \vert - p_c \right)$, where $p_c$ is the momentum at which the exact amplitude (\ref{wfgauge}) of the gauge fields is peaked. Each diagram is therefore proportional to 4 $\delta$-functions, reducing the number of integration variables, and allowing for a reliable evaluation of the integral. 

\begin{figure}[tbp]
\centering 
\includegraphics[width=0.85\textwidth,angle=0]{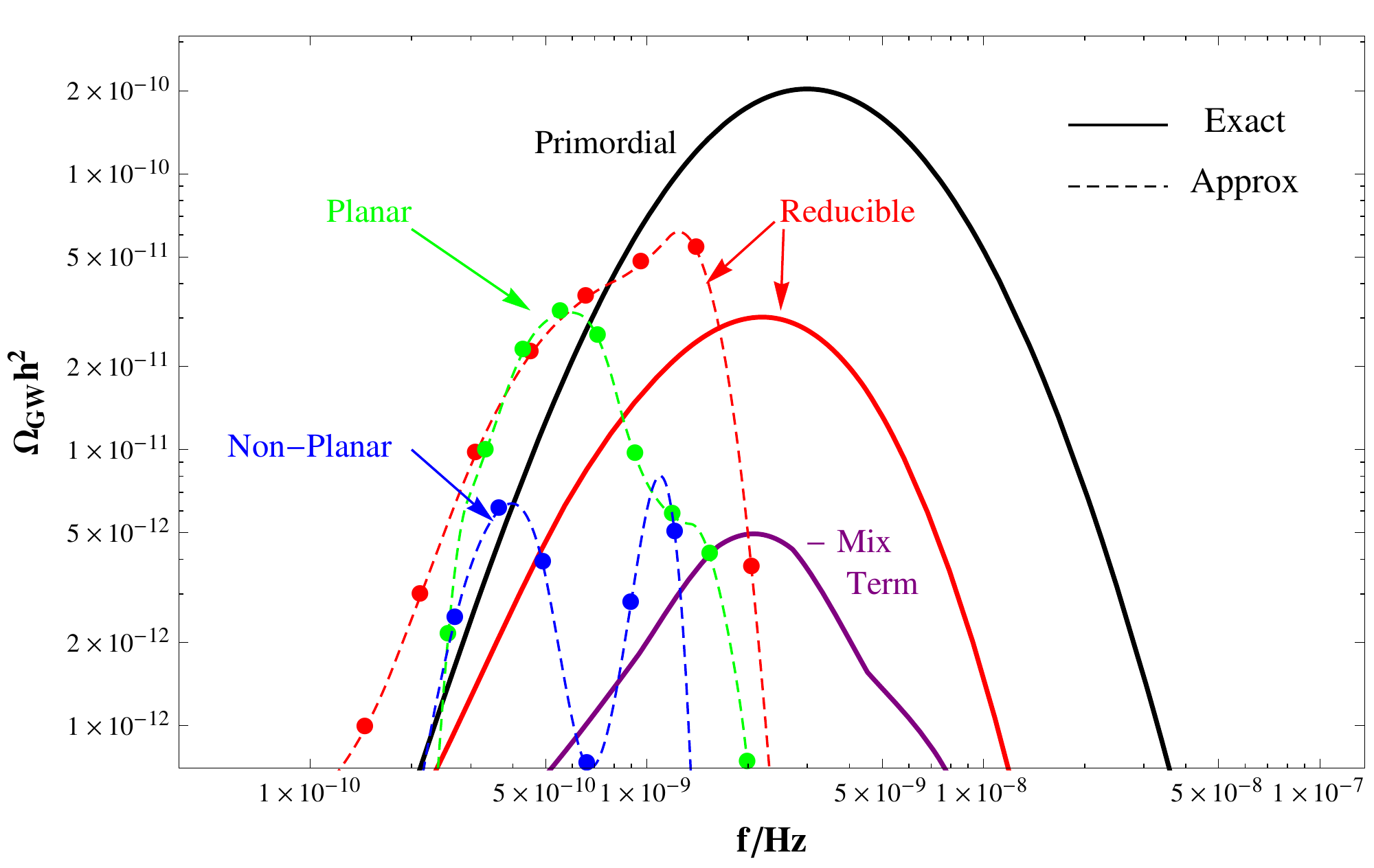}
\hfill
\caption{Comparison of the various contributions to the total GW power spectrum in the rolling axion bump model. Solid lines indicate exact results, obtained using gauge fields mode functions (\ref{wfgauge}). Dashed lines indicate approximate results, obtained using a  ``zero-width'' approximation of the gauge mode functions (the dots indicate the values at which the computation has been performed, and the dashed lines interpolate between these values). 
}
\label{fig:primvsinduced}
\end{figure}

The results of these diagrams are presented,\footnote{The background inflationary evolution assumed in the computation leading to these results is characterized by the slow roll parameters $\epsilon = 6.25\times10^{-4}$ (giving the tensor-to-scalar ratio $r =0.01$ at CMB scales) and $\eta = -0.015$ (giving the Planck \cite{Ade:2015lrj} central value $n_s \simeq 0.965$). We then choose $\delta = 0.2$ in the axion potential, and $\xi_* = 5.1$ (see Appendix \ref{sec:model}).  The axion evolution is chosen so that the axion acquires maximum speed at about $40$ $e$-folds before the end of inflation, producing a peaked GW signal at PTA frequencies. 
The main conclusions of this study (namely, the hierarchy between the various GW signals shown in Figure \ref{fig:primvsinduced}) are unchanged if the spectrum is peaked at different scales, as we discuss at the end of this section.} 
in the red, green, and blue lines respectively of Figure \ref{fig:primvsinduced}. For the Reducible diagram, we show both the exact (obtained using the exact gauge field modes) and the approximated (using the  ``zero-width'' approximation) result. This allows one to quantify the goodness of the approximation: the signal has a peak which is about $2$ times  greater than the exact one,  and centered at a value of $k$ which is about  $2$ times  smaller than the exact one. We therefore expect the approximate results to provide a reliable estimate of the exact ones, up to order one factors. 
 
The total GW spectrum is given by the sum of the auto-correlator $P_{h_i}$, that we just discussed, the auto-correlator  $P_{h_p}$, given in Eq.~(\ref{eq:peakedGW}), and the cross-correlation between the $+$ polarizations of both primordial and induced waves 
\begin{eqnarray} 
P_{h}^{ \, p i }  (\eta, k) &\equiv&  \frac{k^3}{2 \pi^2} \,  \left \langle h_{i,+}  ( {\bf k}) \,  h_{p,+} ( - {\bf k}) 
+   h_{p,+}  ( {\bf k}) \,  h_{i,+} ( - {\bf k})  \right \rangle'  \nonumber\\ 
&& \!\!\!\!\!\!\!\!  \!\!\!\!\!\!\!\!  \!\!\!\!\!\!\!\! 
= \frac{16}{9} \frac{1}{ 2\pi^2} \frac{k^2}{\eta} \int \frac{d^3 p }{(2\pi)^{3/2}} \; p^2 \; \frac{ \sin^2 \theta_{\rm  kp} }{2} \; \int_0^{\eta} d \eta'   \eta' \;  \sin(k\eta - k\eta') \,   F_T ( p \, \eta', \vert {\bf k-p} \vert\ \eta' )  \left  \langle {\hat \zeta}_{\bf p} \,\, {\hat \zeta}_{\bf k-p}\,\, {\hat h}_{p, {\bf k'}}  \right \rangle' \,. \nonumber\\ 
\label{eq:primindgw}
\end{eqnarray} 
The mixed-term  corresponds to the second and third diagram of Figure \ref{fig:hphp-hphi}, and it is evaluated in  Appendix \ref{appsec:cross}, from a two-loop computation that uses the exact expression (\ref{wfgauge}) for the gauge field modes). The result is also shown in Figure \ref{fig:primvsinduced} (where we show it multiplied by $-1$, as the cross-correlation is negative, and the plot has logarithmic axes). 
As can be seen from the figure, the total GW power spectrum is dominated by the primordial one. The Reducible and Planar diagrams provide contributions comparable to each other, which are about $10\%$ of the primordial power spectrum. The Non-Planar diagram, and the cross-correlation term, are further suppressed.

As mentioned above, the  ``zero-width'' approximation squeezes the GW spectrum at smaller frequencies than the exact one, due to the fact that, if we assume that all gauge field momenta are precisely $p_c$, it provides a sharp cut-off on the maximum possible frequency of the produced GW. Therefore, we cannot use this approximation to infer the precise spectrum of the total GW signal, and the broadening of the total GW signal at small frequencies visible in Figure \ref{fig:primvsinduced} should be understood as an artifact of the approximation. We also note that the log-normal shape (\ref{eq:peakedGW}) of the primordial GW power spectrum is a fit which is extremely good at the peak of the sourced signal, but which does not appropriately fit the tail of the bump  \cite{Namba:2015gja}. We verified that the IR tail of the bump scales as $k^3$. 

More in general, the ratio between the induced and the primordial GW PS is an increasing function of the peak amplitude of the gauge field modes, $\frac{P_{h_i}}{P_{h_p}} \propto  \frac{\left\langle \zeta^4 \right\rangle}{\left\langle A^4 \right\rangle} \propto \frac{\left\langle A^8 \right\rangle}{\left\langle A^4 \right\rangle}$. This amplitude grows exponentially with the parameter $\xi_*$ (see Appendix \ref{sec:model}). A growing gauge field amplitude also increases the primordial scalar perturbations, $P_\zeta \propto \left\langle A^4 \right\rangle$, which is limited by the PBH bounds given in Figure \ref{fig:PSZgauschilimits}. In Figure \ref{fig:primvsinduced} we chose $\xi_* = 5.1$, which is the largest value allowed by the PBH limits on  $\xi_*$ for a bump at the chosen scale. We find that $\frac{P_{h_i}}{P_{h_p}}$ could be of order one only for an increase in gauge field production that would lead to an increase of the scalar power spectrum by about a factor of $10$. This would violate the PBH bounds at all scales shown in Figure \ref{fig:PSZgauschilimits}. We conclude that in the rolling axion bump model, the primordial GW always dominate over the induced ones.~\footnote{While  in this section we have fixed  $\delta = 0.2$, in the following sections we consider cases with greater $\delta$. As seen in \cite{Namba:2015gja} the ratio between primordial GW and primordial scalar perturbations is an increasing function of $\delta$. Therefore the induced GW are subdominant in all the rolling axion cases studied in this paper.}

%%%%%%%%%%%%%%%%%%%%%%%%%%%%%%%%%%%%%
\section{Stochastic GW spectra as a probe of PBH dark matter  }
\label{sec:gaussvsnongauss}
%%%%%%%%%%%%%%%%%%%%%%%%%%%%%%%%%%%%%

We now use the results of the previous section to understand what can be learned from the observation of a bump of primordial and induced GW associated with the PBH. We divide the discussion into two Subsections, devoted to the study of a signal at PTA and at LISA scales, respectively.

%%%%%%%%%%%%%%%%%%%%%%%%%%%%%%%%%%%%%
\subsection{GW at PTA scales} 
\label{sec:PTA}
%%%%%%%%%%%%%%%%%%%%%%%%%%%%%%%%%%%%%

PTA measurements are most sensitive at frequencies $f \sim$ few~nHz. GW modes of such frequencies originate from scales that left the horizon about $N \sim 40$ $e$-folds before the end of inflation (as can be immediately seen by combining Eqs.~(\ref{M-f}) and (\ref{M-N})). From Eq.~(\ref{M-N}), we see that scalar overdensities produced at $N \sim 40$ $e$-folds collapse into primordial black holes of mass $M \sim {\cal O } \left( 10 \right) M_\odot$. Therefore, as already pointed out in \cite{Garcia-Bellido:2016dkw}, PTA measurements can provide useful information on PBH of such masses.

\begin{figure}[tbp]
\centering 
\includegraphics[width=0.48\textwidth,angle=0]{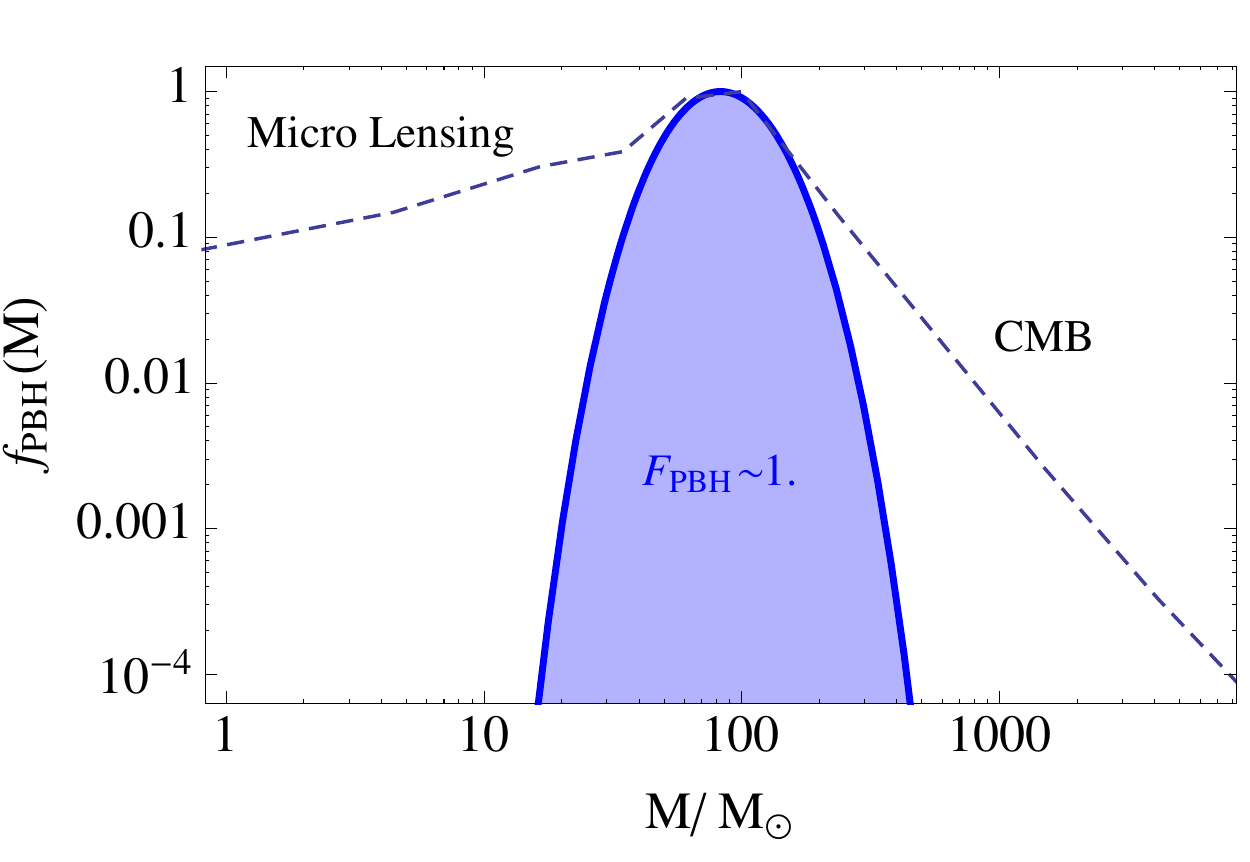}
\hfill
\caption{Example of a distribution of PBH masses that satisfies the current PBH bounds and that can account for the present dark matter of the universe, $F_{\rm PBH} \equiv \int dM/M \, f_{\rm PBH} \left( M \right) = 1 $.  The  distribution has a shape as obtained from the rolling axion bump model  \cite{Namba:2015gja,Garcia-Bellido:2016dkw}, which is well approximated by a log-normal distribution close to the peak, $f_{\rm PBH} \left( M \right) = \frac{1}{\sqrt{2 \pi} \, \sigma} \, {\rm exp} \left( - (\ln M / M_{\rm peak})^2/(2 \sigma^2) \right)$.  The distribution shown in the Figure is characterized by $M \simeq 83 M_\odot$ and $\sigma \simeq 0.42$, in a window where the PBH bounds are weakest. This distribution is used in producing the GW signals shown in Figs.~\ref{fig:G-NG} and \ref{fig:PTAevol}. 
}
\label{fig:f(M)PTA}
\end{figure}

\begin{figure}[tbp]
\centering 
\includegraphics[width=0.48\textwidth,angle=0]{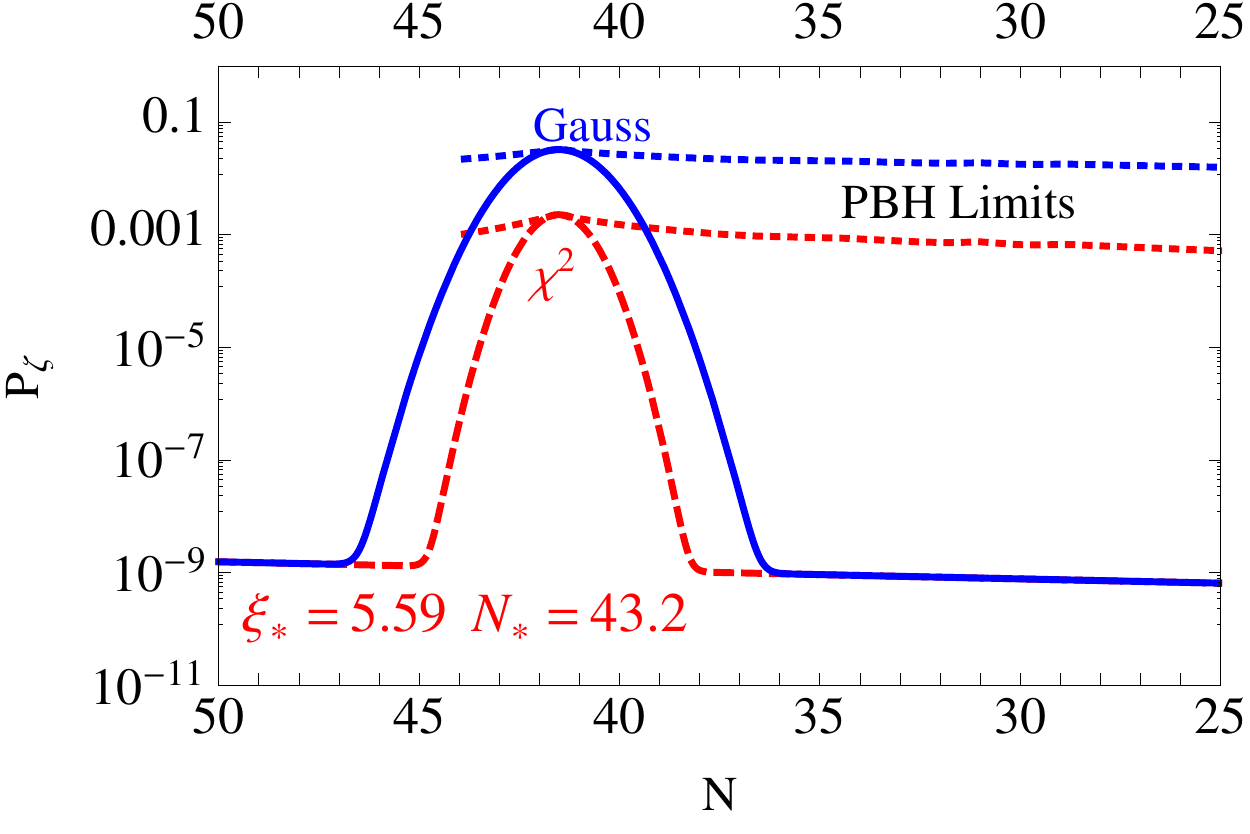}
\includegraphics[width=0.48\textwidth,angle=0]{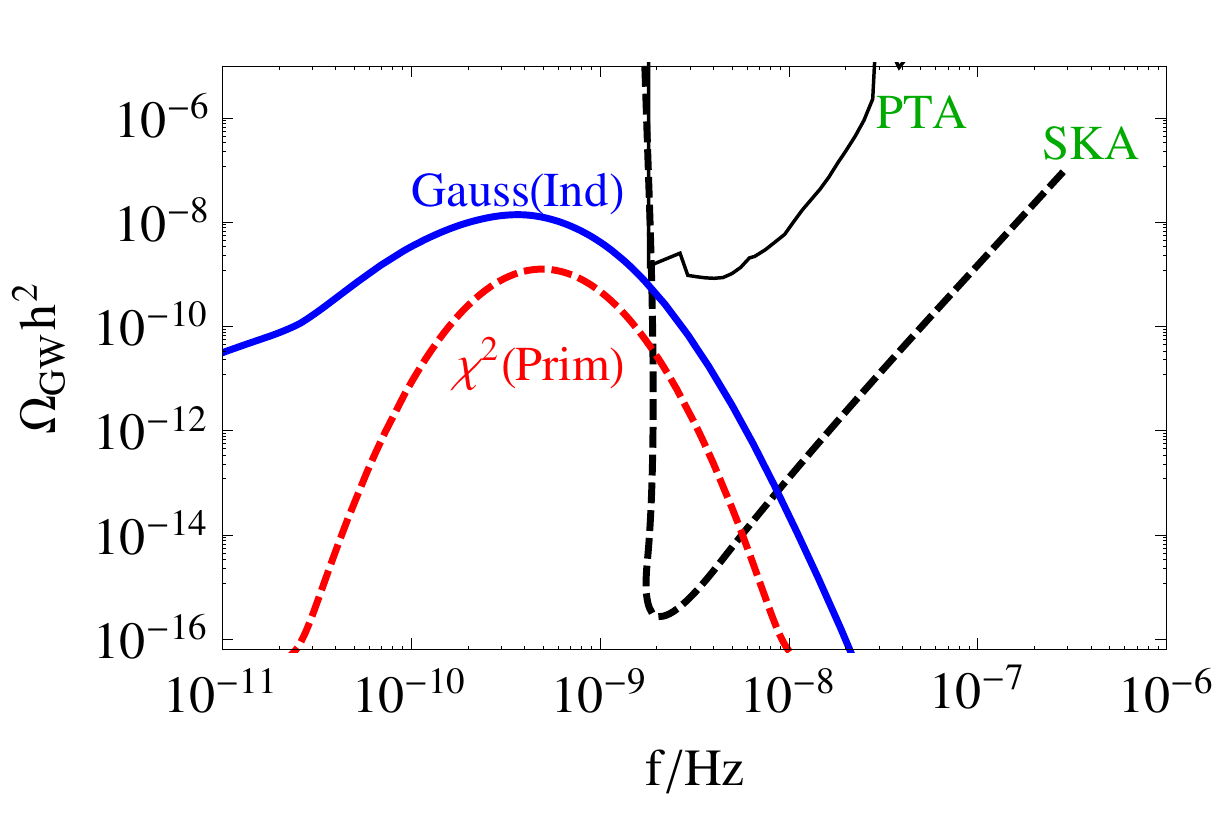}
\hfill
\caption{Left panel: Bump in the primordial scalar perturbations that saturates the PBH bound at PTA scales. 
Right panel: Corresponding bump in the stochastic GW background, compared with current (``PTA'') and forthcoming (``SKA'') limits. 
The blue solid (resp., red dashed) curves refer to the Gaussian bump model (resp., the rolling axion bump model, for which we chose 
$\xi_* = 5.59$ and $\delta = 0.4$). 
}
\label{fig:G-NG}
\end{figure}

We quantify this in the context of the Gaussian vs. Non-Gaussian (rolling axion) bump models studied in the previous sections. 
In Figure~\ref{fig:f(M)PTA} we show  a distribution of current PBH masses,  that  saturates the PBH limit in this mass range and that constitutes all of the dark matter of the universe, 
\begin{equation}
F_{\rm PBH} \equiv \int \frac{dM}{M} \, f_{\rm PBH} \left( M \right) = 1 \;. 
\end{equation} 
In the left panel of Figure \ref{fig:G-NG} we show the bump in the primordial scalar curvature required to produce this distribution, both in the case of the Gaussian peak and of the rolling axion peak models.\footnote{We stress that this discussion ignores any possible merging and accretion of the PBH after their formation. For a proper discussion, see the next section.} We note that the required distribution of $P_\zeta$ in the Non-Gaussian case is much smaller, and narrower, than the required distribution in the Gaussian case. Nevertheless, they result in the same $f_{\rm PBH} \left( M \right)$, due to the very different relations (\ref{gauss-chi2}) for the PBH formation fraction $\beta$. 

 In both models, this bump in the scalar modes is accompanied by a GW bump at PTA frequencies. In the Gaussian bump model, the GW signal is sourced by the scalar perturbations at horizon re-entry. In the Non-Gaussian rolling axion bump model (denoted as $\chi^2$ in the figure), the GW signal is dominated by the primordial GWB produced during inflation, by the same mechanism that produced the bump in the scalar modes. As we already discussed in the previous section, we stress that the induced GW signal is much smaller in the Non-Gaussian vs. the Gaussian model, since the PBH bound on the scalar perturbations is much more stringent in the former case (a more constrained $\zeta$ implies a more constrained induced $\zeta + \zeta \rightarrow h_i$ signal). 

The magnitude of this GW signal is shown in the right panel of Figure \ref{fig:G-NG}, where it is compared with the present PTA bounds~\cite{Arzoumanian:2015liz,Lentati:2015qwp,Shannon:2015ect}, as well as the forecast bounds for the forthcoming Square Kilometer Array (SKA) experiment  \cite{Moore:2014lga,Zhao:2013bba}.  While consistent with the current bounds, both models produce a GW signal  well within the reach of SKA.

Besides the PBH limit shown in Figure \ref{fig:PSZgauschilimits}, the spatial curvature perturbations are also constrained by $\mu$ and $y$ CMB distortions. Of relevance for the present discussion, see also Ref.~\cite{Clesse:2015wea}, the $\mu$ distortion is  given by \cite{Chluba:2015bqa,Nakama:2017ohe}
\begin{eqnarray} 
\!\!\!\!\!\!\!\! \!\!\!\! 
\mu  \simeq  -3 \times 10^{-9}  +  2.3 \int_{k_0}^\infty \frac{dk}{k} P_\zeta \left( k \right)  \left[ {\rm exp} \left( - \frac{\left[ \frac{\hat k}{1360} \right]^2}{1+ \left[ \frac{\hat k}{260} \right]^{0.3} + \frac{\hat k}{340}} \right) -  {\rm exp} \left( - \left[ \frac{\hat k}{32} \right]^2 \right) \right] , 
\end{eqnarray} 
where ${\hat k} = k \, {\rm Mpc}$ and ${\hat k}_0 =1$. In this expression, the primodrial curvature power spectrum is multiplied  by a window function with its main support at wavenumbers $50 \la \hat k \la 2\times10^4$. Assuming $N_{\rm CMB} = 60$ at the scale $\hat k_{\rm CMB}=0.002$, this corresponds to modes that left the horizon between approximately $45$ and $50$ e-folds before the end of inflation.\footnote{On the other hand, $y$-distortions are mostly sensitive to modes $1 \la \hat k \la 50 $, which roughly corresponds to $50 \la N \la 54$. These scales are significantly larger than those considered in this work.}  The Gaussian and $\chi^2$ distribution shown in Figure \ref{fig:G-NG} lead, respectively, to the distortion $\mu \simeq 3.6 \times 10^{-5}$ and  $3 \times 10^{-8}$. Both values are below the current bound  $\vert \mu \vert \la 10^{-4}$ from the COBE / FIRAS experiment~\cite{Mather:1993ij,Fixsen:1996nj}. The CMB distortion obtained in the Gaussian bump model is well within the reach of a  PIXIE-like experiment, which has an estimated sensitivity  $\vert \mu \vert = {\cal O} \left( 10^{-7} \right)$~\cite{Abitbol:2017vwa}. The rolling axion model leads instead to a value below this sensitivity, and only slightly greater than the scale invariant case (a scale invariant spectrum corresponding to that of Figure \ref{fig:G-NG}, with no bump, leads to  $\mu \sim 10^{-8}$).

We also see from the figure that the Gaussian bump model results in a much greater GW signal than the rolling axion bump model, and that the Gaussian bump case shown in the figure is only barely compatible with the present bounds. It is therefore important to understand how this conclusion is affected if we modify the PBH distribution with respect to the one shown in Figure \ref{fig:f(M)PTA}. The most important  factor in this discussion is the relation between the peak frequency of the GW signal and the peak mass of the PBH distribution. From Eq.~(\ref{M-k}) we see that the PBH mass $M$ is related to the peak frequency of the scalar perturbations ($f = k/2\pi$) by   
\begin{equation}
M \simeq 50 \,  \gamma\,  M_\odot   \, \left(  \frac{10^{-9} \, {\rm Hz}}{f} \right)^2  \,.
\label{M-f}
\end{equation}
The peak frequency in the scalar and GW signal are equal to each other, up to an order one factor. Therefore, the peak frequency of the GW signal potentially detectable in PTA experiments scales with the peak mass of the PBH distribution as $f_{\rm peak} \propto M_{\rm peak}^{-1/2}$.

\begin{figure}[tbp]
\centering 
\includegraphics[width=0.48\textwidth,angle=0]{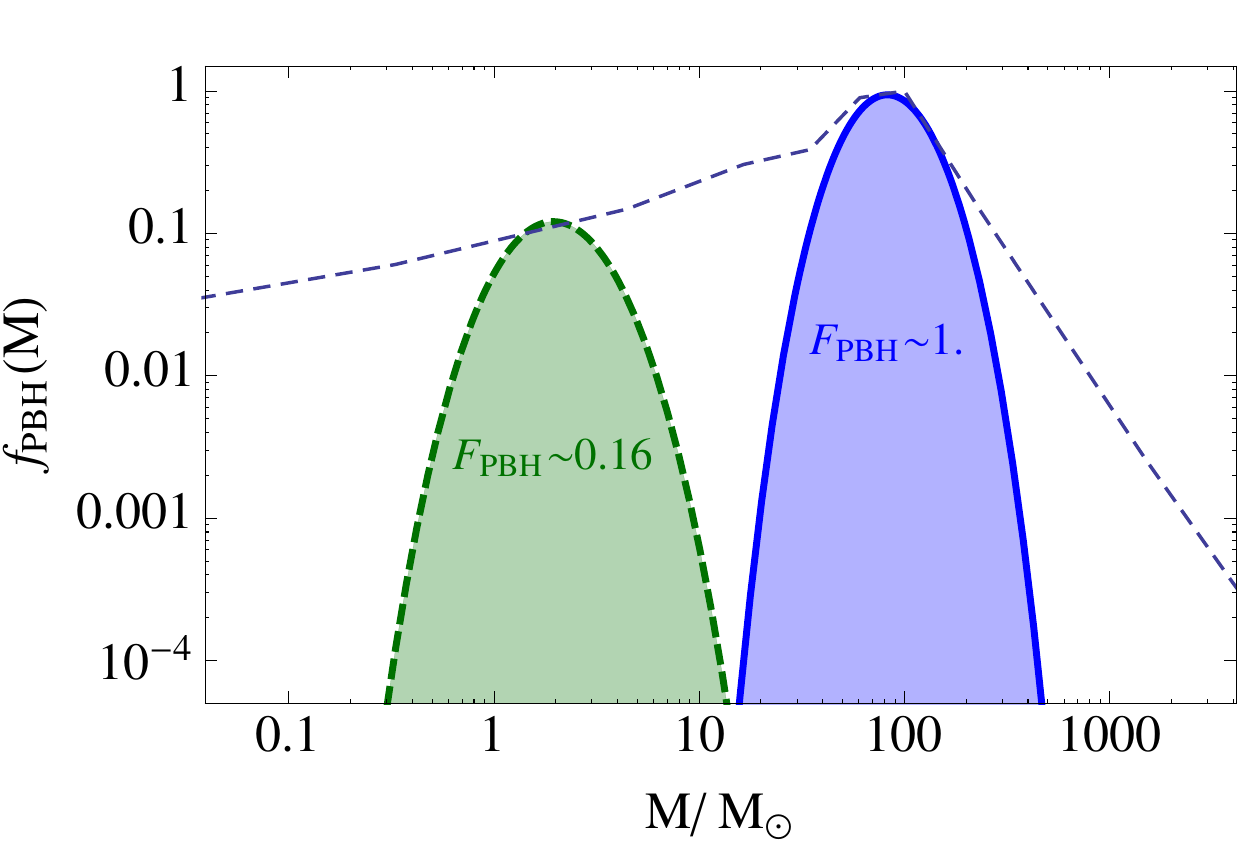}
\includegraphics[width=0.48\textwidth,angle=0]{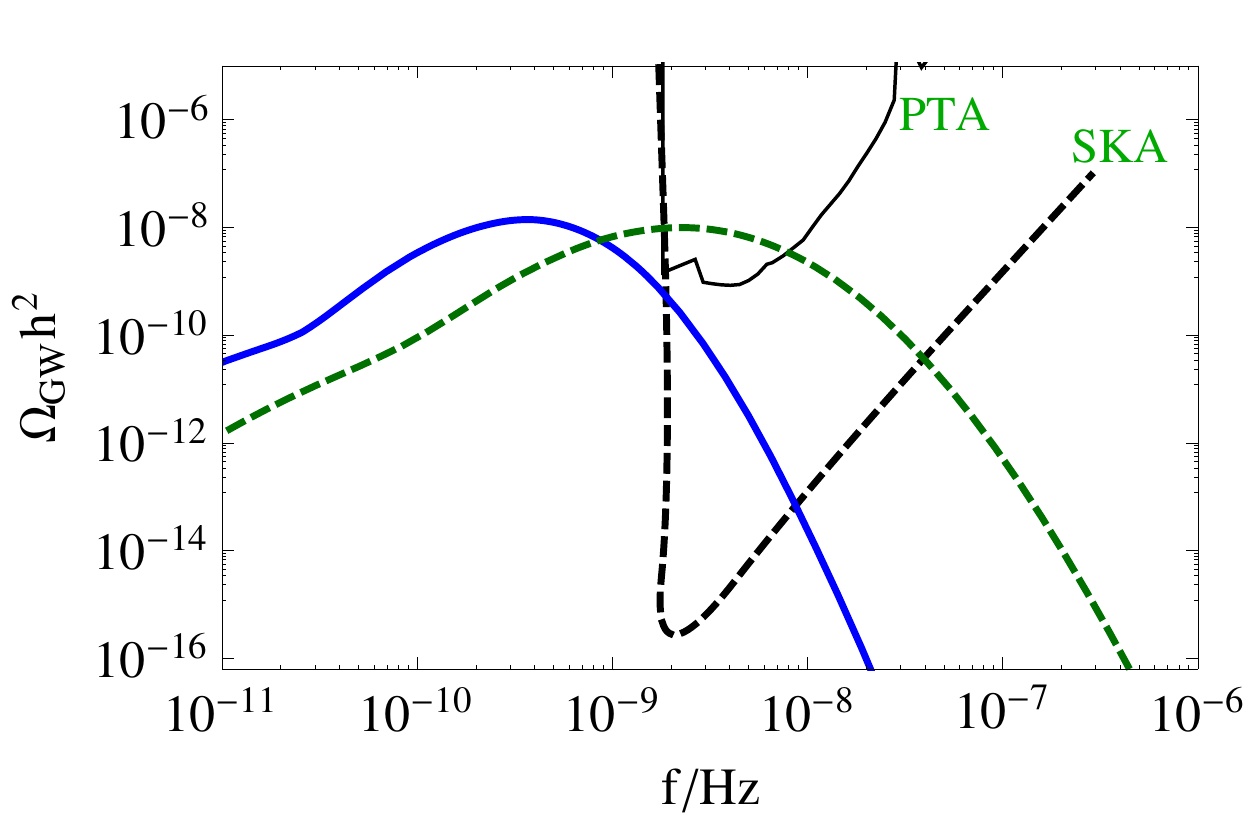}
\hfill
\caption{Dependence of the PTA constraints on the PBH mass distribution. The PBH mass distribution shown in Figure \ref{fig:f(M)PTA} (in blue solid lines) is compared here with a distribution peaked at smaller PBH mass (in green dashed lines).   The GW signal produced by this second distribution is ruled out by the present PTA data. Both models shown in this figure assume a Gaussian statistics. 
}
\label{fig:G-compare}
\end{figure}

In Figure~\ref{fig:G-compare}  we compare the GW signal generated by the Gaussian bump model described above with that generated by a different Gaussian bump model, peaked at a smaller PBH mass. More precisely, the second model (shown in dashed green lines in Figure~\ref{fig:G-compare}) is peaked at  $M \simeq 2 \, M_\odot$, a factor $\sim 41$ smaller than the peak value  $M \simeq 83 \, M_\odot$, of the first model (shown in solid blue lines). The second model produces a GW signal peaked at $f \sim 2.3 \, {\rm nHz}$, in a region where the PTA bounds are strongest. This frequency is a factor $\sim 6.2$ greater than the peak frequency $f \sim  0.37 \, {\rm nHz}$ of the GW signal generated in the first model, in very good agreement with the  $f_{\rm peak} \propto M_{\rm peak}^{-1/2}$ scaling. Despite the fact that the second PBH distribution only accounts for $\sim 16\%$ of the dark matter of the present Universe, the shift in frequency causes it to be already ruled out by the PTA data.

It is also important to stress that the examples studied in Figures~\ref{fig:G-NG} and \ref{fig:G-compare} assume $\zeta_c = 1$ in Eq.~(\ref{gauss-chi2}). This quantity is the estimated threshold that a scalar perturbation must reach in order to form a PBH. Theoretical and numerical studies~\cite{Carr:1975qj,Nakama:2013ica,Shibata:1999zs,Polnarev:2006aa,Khlopov:2008qy} indicate that this quantity is $\zeta_c = {\cal O } \left( 0.05 - 1 \right)$. Since the amount of PBH is controlled by the ratio $\sqrt{P_\zeta}/ \zeta_c$, a decrease of $\zeta_c$ by a factor $r$ leads to the same PBH abundance provided that $P_\zeta$ is decreased by $r^2$. This effect decreases the GW signal by $r^2$ in the rolling axion bump model (in which both $P_\zeta$ and $P_{\rm GW}$ are proportional to the same power of the sourcing fields), and by $r^4$ in the Gaussian model (in which $P_{\rm GW} \propto P_\zeta^2$). Therefore, for all values of 
 $\zeta_c = {\cal O } \left( 0.05 - 1 \right)$, the GW produced by these models will be testable at PTA-SKA frequencies. 

We conclude that a significant dark matter component in the form of PBH with masses in the range $M \sim 1 - 100\, M_\odot$ is compatible with the current PTA bounds for the rolling axion bump model, and barely compatible or excluded for the Gaussian bump model, depending on the precise peak PBH mass and on the value of the threshold parameter $\zeta_c$.  The forthcoming improvement of several orders of magnitude on the PTA bounds  from the SKA experiment will allow to conclusively probe both models. In Section \ref{sec:evol} we discuss how this conclusion is modified by a nontrivial evolution (via accretion and merging) of the PBH distribution after their formation.

%%%%%%%%%%%%%%%%%%%%%%%%%%%%%%%%%%%%%
\subsection{GW at LISA scales} 
\label{sec:LISA}
%%%%%%%%%%%%%%%%%%%%%%%%%%%%%%%%%%%%%

Here we study the implications of LISA measurements on the PBH physics. The LISA experiment will be most sensitive at frequencies $f \sim {\rm few} \ m$Hz, see Ref.~\cite{Bartolo:2016ami}. This corresponds to modes that left the horizon about $N \sim 25$ $e$-folds before the end of inflation. From Eq.~(\ref{M-N}), we see that scalar overdensities produced at $N \sim 25$ collapse into primordial black holes of mass $M \simeq {\rm few} \times 10^{-12} \, M_\odot$. Therefore, LISA measurements can provide useful information on PBH of such small masses. 

\begin{figure}[tbp]
\centering 
\includegraphics[width=0.48\textwidth,angle=0]{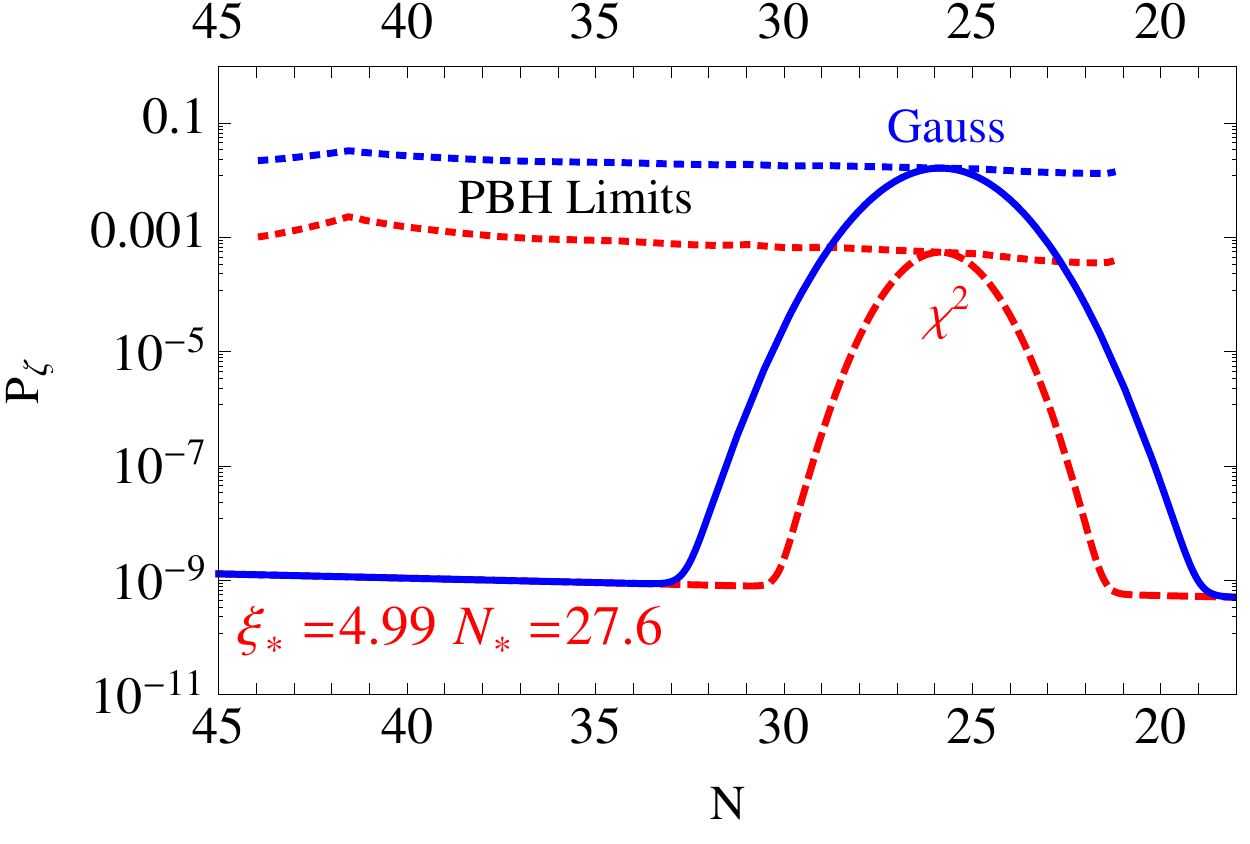}
\includegraphics[width=0.48\textwidth,angle=0]{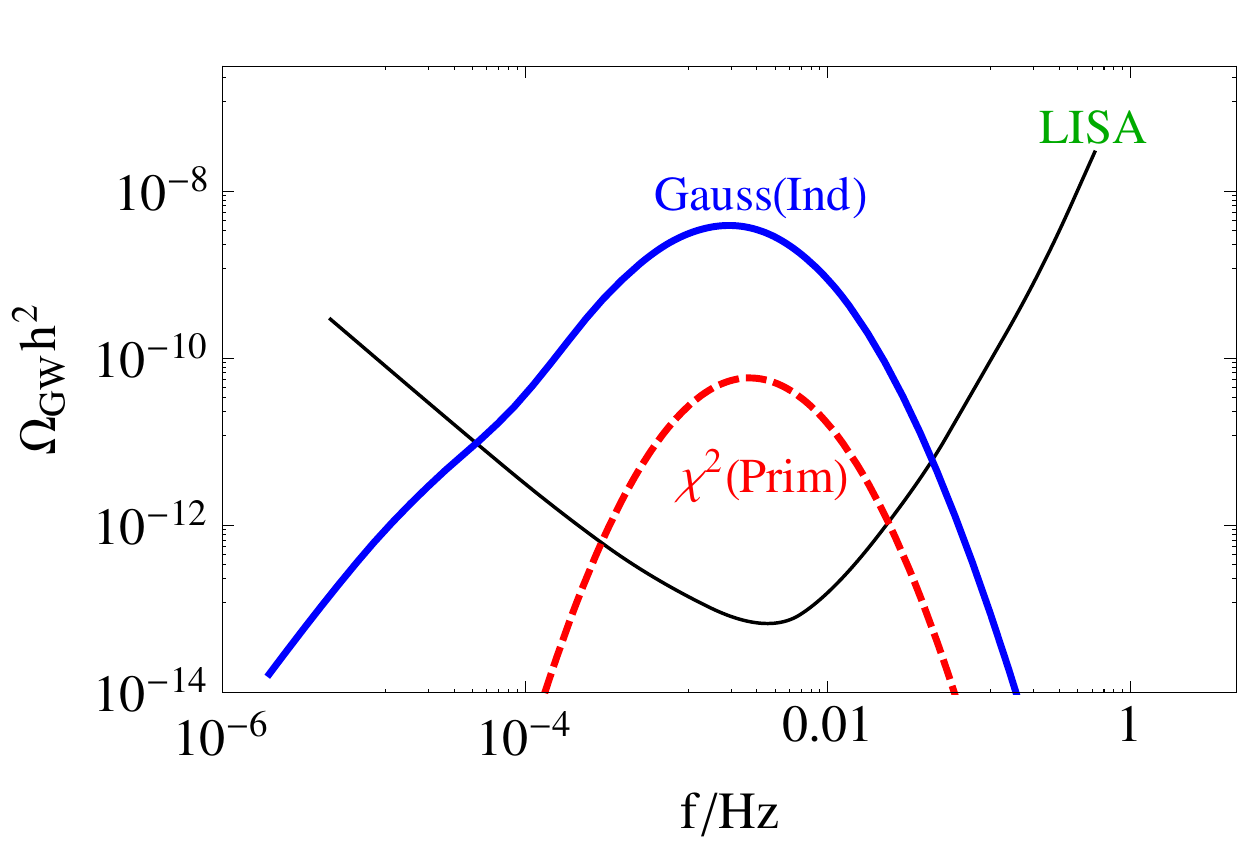}
\hfill
\caption{
Left panel: Bump in the primordial scalar perturbations that saturates the PBH bound at LISA scales. 
Right panel: Corresponding bump in the stochastic GW background. The blue solid (resp., red dashed) curves refer to the Gaussian bump model (resp., the rolling axion bump model). 
}
\label{fig:G-NGLISA}
\end{figure}

Analogously to the previous subsection, in the left panel of Figure~\ref{fig:G-NGLISA} we show a bump in the primordial curvature perturbations that saturates the present PBH bounds, given by neutron star capture \cite{Carr:2016drx}. The curves shown in the Figure correspond to a present PBH dark matter fraction equal to one (this mass range was also recently considered in Ref.~\cite{Inomata:2017okj}).  In the right panel of Figure~\ref{fig:G-NGLISA} we show the corresponding bump in the GW spectrum, as compared with the  forecasted LISA sensitivity curve ``N2A2M5L6''~\footnote{This is the sensitivity curve, among those considered in Ref.~\cite{Bartolo:2016ami}, that is expected to be the closest to the final LISA configuration. We thank Chiara Caprini for discussions.}  given in Ref.~\cite{Bartolo:2016ami}.

As seen from the right panel, the GW signal from the Gaussian model (resp., from the rolling axion model) is about five orders of magnitude (resp., three orders of magnitude) stronger than the best sensitivity curve of LISA. As discussed in the previous subsection, the GW signal can be decreased, while keeping the same amount of PBH, if the threshold for formation $\zeta_c$ is lowered with respect to the value $\zeta_c = 1$ assumed in Figure \ref{fig:G-NGLISA}. We find that the GW signal is below the LISA sensitive curve only for $\zeta_c \la 0.07$ in the Gaussian bump model, and for $\zeta_c \la 0.03$ in the rolling axion bump model. Therefore, if PBH in the mass range $M\sim10^{-12} M_\odot$ constitute a significant fraction of the dark matter, the associated stochastic background of primordial or induced GW produced by the bump models will be observed by LISA.\footnote{The star formation limits of~\cite{Capela:2014ita} are compatible with a PBH dark matter fraction of ${\cal O } \left( 10^{-4} \right)$. The bounds on $P_\zeta$ shown in Figure \ref{fig:PSZgauschilimits} change very slowly with $f_{\rm PBH}$ (as can be understood from Eq.~(\ref{gauss-chi2}). See for instance Figure 9 of \cite{Garcia-Bellido:2016dkw}). As a consequence, even $F_{\rm PBH} = {\cal O } \left( 10^{-4} \right)$ would result in a visible GW signal at LISA.}

%%%%%%%%%%%%%%%%%%%%%%%%%%%%%
\section{Evolution of the PBH Mass Function }
\label{sec:evol}
%%%%%%%%%%%%%%%%%%%%%%%%%%%%%

In this section we discuss the effect of accretion and merging on the PBH mass function, and how they can impact the observation at PTA scales of the primordial and induced GW signal studied in the previous sections. In appendix \ref{appsec:merging} we obtained the present time ($t_0$) PBH fraction 
\begin{equation} 
f_{\rm PBH} \left( M ,\, t_0 ,\, {\cal A} ,\, {\cal M} \right) \simeq  6.7 \times 10^8 \,  \gamma^{1/2}  {\cal A} \, \sqrt{ \frac{{\cal A} \, {\cal M} \, M_{\odot}}{{M}}} \; \beta \left( \frac{M}{{\cal A}\,{\cal M}\,M_\odot} \right) \;. 
\label{f-A-M}
\end{equation} 
This relation was derived in \cite{Carr:2009jm} in the case of no accretion (${\cal A} = 1$) and no merging (${\cal M} = 1$) of the PBH after their formation. The coefficient ${\cal A}$ in  (\ref{f-A-M}) accounts for the fact that the surrounding plasma can accrete a PBH after its formation. We parametrize this by an increase of each individual PBH mass, $M_f \rightarrow M = {\cal A} \, M_f$, from its formation value $M_f$ to its present value $M$, and by an overall increase of the total energy density in PBH (see Eq.~(\ref{F-A})).  The coefficient ${\cal M}$ instead accounts for the shift of the PBH distribution towards greater mass due to merging of individual PBH,  $M_f \rightarrow  {\cal M} \, M_f$, without increasing the overall energy density. As a combined effect, the peak of the PBH distribution moves according to 
\begin{equation}
M_f \rightarrow M = {\cal A} \, {\cal M} \, M_f \,, 
\label{shift-M}
\end{equation}
so that a distribution that was originally a function of $M_f$ becomes a function of $M/{\cal A} \, {\cal M}$, while the overall integral $f_{\rm PBH} = \int dM/M \, f_{\rm PBH}(M) = \rho_{\rm PBH}/\rho_{\rm CDM}$ increases by the factor~${\cal A}$. 

This rough parametrization does not account for the fact that, in reality, different PBH masses will merge and accrete differently, and that this will in general change the {\em shape} of the original distribution. This distortion will be  subdominant if the PBH initial distribution is sufficiently peaked (since in this case, only one mass scale is relevant). As we  now discuss, the main impact of  accretion and merging on PTA observations is simply due to a shift of the PBH peak mass (\ref{shift-M}), and the precise shape of the final PBH distribution plays a far less relevant role.

\begin{figure}[tbp]
\centering 
\includegraphics[width=0.48\textwidth,angle=0]{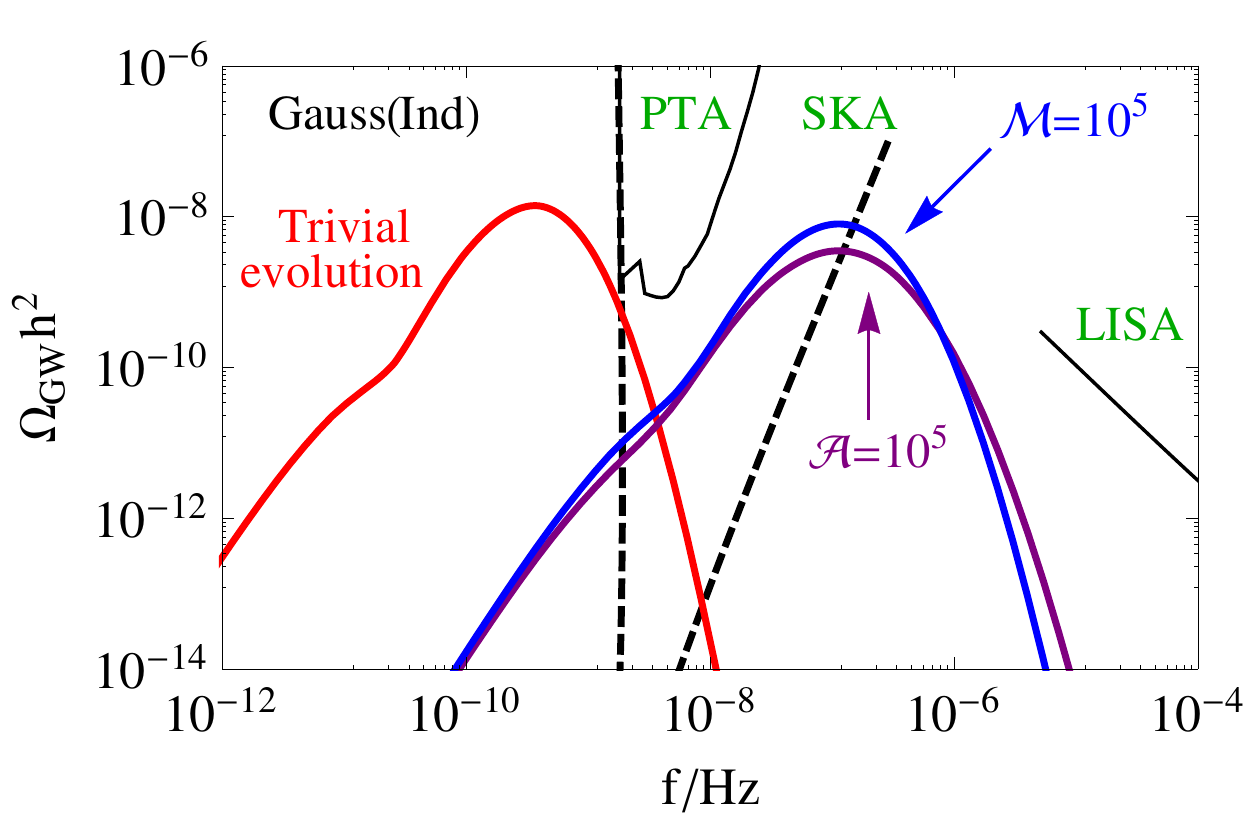}
\includegraphics[width=0.48\textwidth,angle=0]{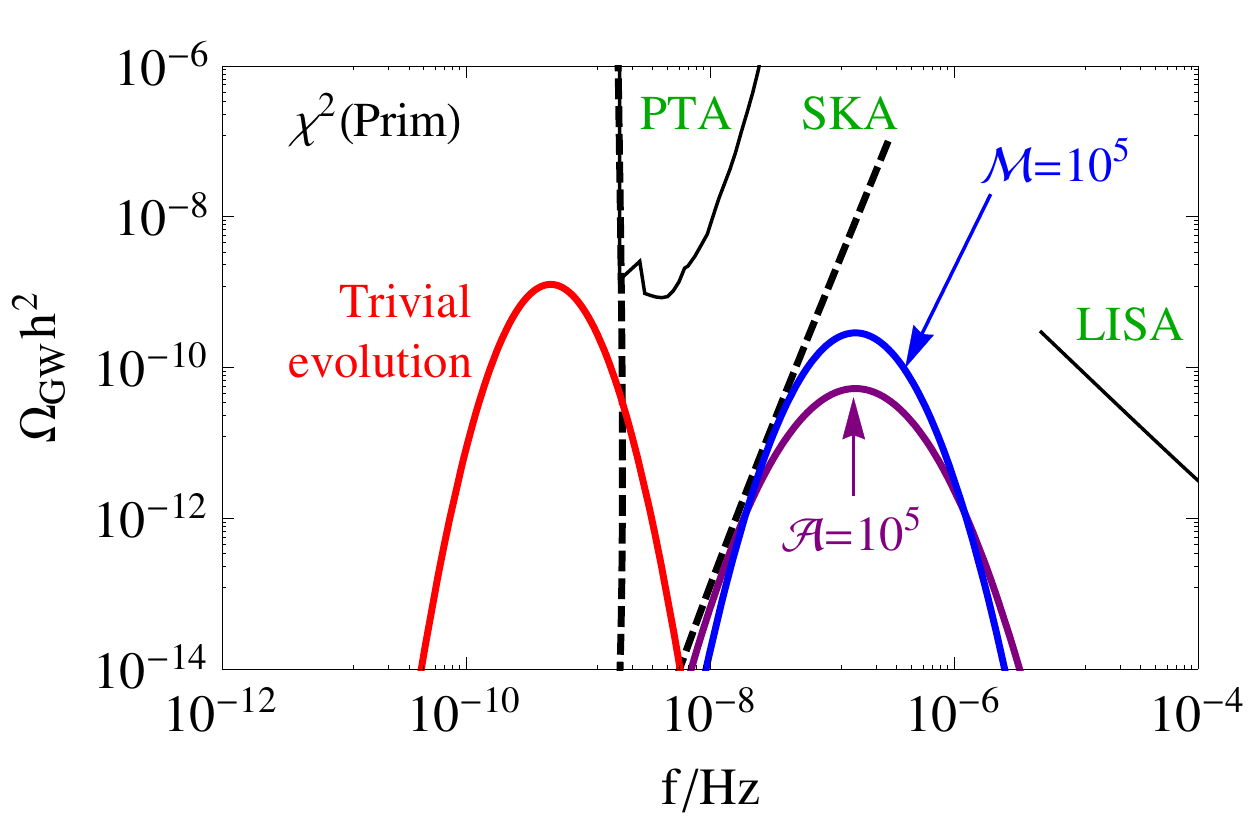}
\caption{Primordial and induced GW obtained from different models that lead to the current PBH distribution shown in Eq.~(\ref{fig:f(M)PTA}). The various models differ for the assumed statistics of the primordial curvature and for the amount of merging (${\cal M}$) and accretion (${\cal A}$) of the black holes after their formation. 
}
\label{fig:PTAevol}
\end{figure}

This can be understood from the examples shown in Figure \ref{fig:PTAevol}. In the left (respectively, right) panel of the figure we assume the Gaussian bump (resp., rolling axion bump) model, namely a Gaussian (resp., $\chi^2$) statistics for the primordial perturbations. In each figure the PTA limits are compared with three GW distributions, obtained in the case of trivial PBH evolution (red curve, ${\cal M} = {\cal A} = 1$), of very strong merging and no accretion (blue curve,  ${\cal M} = 10^5$ and ${\cal A} = 1$), and of very strong accretion and no merging (purple curve,  ${\cal M} = 1$ and ${\cal A} = 10^5$). We expect that realistic amounts of accretion and merging should lie between these extreme cases. The GW signals are obtained as follows. In all cases we assume a present PBH mass function given by Figure \ref{fig:f(M)PTA}. This present distribution is peaked at $M \sim 83 \, M_\odot$, and it accounts for all the dark matter of the universe. We then find the corresponding value for the PBH fraction at formation $\beta$, according to Eq.~(\ref{f-A-M}), accounting for the different values of ${\cal M}$ and ${\cal A}$ that characterize each case. We then compute the primordial curvature power spectrum $P_\zeta$ leading to this fraction, in the two different cases of Gaussian vs. $\chi^2$ distribution. Finally, we compute the corresponding amount of primordial and induced GW associated with this distribution. The various GW signals obtained in this way are plotted in the various curves of Figure \ref{fig:PTAevol}.~\footnote{We find that the width of the $\chi^2$ distributions are $\delta \simeq 0.4$ in the case of trivial evolution, $\delta \simeq 0.3$ in the case of strong merging, and  $\delta \simeq 0.2$ in the case of strong accretion. The width of the each gaussian bump is given by  the width of the corresponding $\chi^2$ distribution times $\sqrt{2}$. }

The main feature that emerges when comparing the merging or accretion cases with the trivial evolution case is the increase of the peak frequency of the GW signal. The reason is the following: the distribution (\ref{fig:f(M)PTA}) probes the current PBH mass $M$. On the contrary, the primordial and induced GW signals shown in Figure \ref{fig:PTAevol} depend on the PBH formation mass, $M_f = M/{\cal A} {\cal M}$. This decreased mass results in an increase of the frequency of the GW signal by $\sqrt{{\cal A} {\cal M}}$, due to the $f \propto M_f^{-1/2}$ scaling discussed after Eq.~(\ref{M-f}). This shift of the GW distribution is the main factor in determining whether the GW signal can be probed at PTA scales. We see that, for the case of a $\chi^2$  bump, an accretion or a merging of a $\sim 10^5$ factor would shift the GW signal towards too high frequencies to be observed in these experiments. The Gaussian model results instead in a greater and wider GW signal, that can be observed at PTA scales also for these large amounts of accretion and merging. 

It is also interesting to compare the signal in the case of accretion and no merging, vs. the case of merging and no accretion. In the case of only accretion, the primordial signal must have a smaller amplitude with respect to the case of only merging. This results in smaller primordial and induced GW, as clearly visible in the figure. The difference in the amplitude of the two signals is however rather limited. This is due to the strong sensitivity of the PBH abundance to small changes of the primordial curvature (see Eq.~(\ref{gauss-chi2}) of the present work, and Figure 9 of  \cite{Garcia-Bellido:2016dkw}). If we want to produce the same PBH abundance today, a $10^5$ accretion requires that the initial PBH abundance is decreased by a $10^5$ factor. This however only requires a small decrease of the primordial curvature spectrum, so that the primordial and induced GW signals decrease by less than one order of magnitude. 

Finally, let us comment on the $\mu$-distortion obtained in the case of large accretion and/or merging. At equal present PBH distributions, large accretion and merging imply a smaller formation mass $M_f$. This in turn implies a shift of the primordial curvature modes to smaller scales~(\ref{M-k}), and a later formation time (smaller number of e-folds $N$) during inflation~(\ref{M-N}). For  the cases of ${\cal M} = 10^5$ and ${\cal A} = 10^5$ studied in this section, we find that, due to this shift, the bump does not induce additional $\mu$-distortions with respect to a scale invariant spectrum.

To conclude, the main effect of merging and accretion on the detectability of the GW signal at PTA scales is related to the fact that they increase the frequency of the GW signal by the collective $\sqrt{{\cal A} {\cal M}}$ factor. For realistic values of merging and accretion, we expect that a significant PBH distribution of ${\cal O} \left( 10 \right) M_\odot$ will be associated to a GW signal visible at PTA-SKA frequencies. Interestingly, the comparison between the peak of the present PBH distribution and the peak of this GW signal will allow us to obtain information on  $\sqrt{{\cal A} {\cal M}}$, and therefore on the evolution of the PBH subsequent to their formation.

%%%%%%%%%%%%%%%%%%%%
\section{Conclusion and Outlook } 
\label{sec:conc}
%%%%%%%%%%%%%%%%%%%%

Cosmic Microwave Background \cite{Ade:2015lrj} and Large Scale Structure (LSS) \cite{Cuesta:2015mqa} measurements strongly  support the paradigm of primordial inflation. However, they only probe the range of wave numbers $10^{-4} \, {\rm  Mpc}^{-1} \la k \la 0.1 \, {\rm Mpc}^{-1}$, corresponding to about $7$ $e$-folds of inflation. We have currently little or no direct experimental information on the physics of inflation at later times / smaller scales, apart from the limits on the amount of scalar perturbations resulting from bounds on  PBH and ultracompact minihalos \cite{Bringmann:2011ut,Emami:2017fiy}. If present, PBH can provide a new experimental window on smaller scales than those relevant for CMB and LSS observations \cite{GarciaBellido:1996qt}. Of particular interest is the $M\sim 1 - 100~M_\odot$ range of PBH masses. PBH in this range may be responsible for the GW signal observed at LIGO~\cite{Bird:2016dcv,Clesse:2016vqa,Sasaki:2016jop}. Moreover, the revision of the CMB bounds on PBH in this mass range \cite{Ali-Haimoud:2016mbv}, and uncertainties on the  Eridanus II limits \cite{Li:2016utv} allow for the possibility that a distribution of PBH masses in this range could constitute all of the dark matter in the Universe. 

This mass interval corresponds to modes that left the horizon about  $40$ $e$-folds before the end of inflation (see Eq.~\ref{M-N}, where we assume that $N_{\rm CMB} = 60$). This scale corresponds to a present frequency of  $f = {\cal O }$(nHz), which can be probed at PTA.\footnote{Another possible range where, in light of the uncertainty of the present experimental limits,  a distribution of PBH masses may give the total dark matter is around $M \sim 10^{-12} M_\odot$. This corresponds to modes produced at $N \sim 25$ $e$-folds before the end of inflation, and to a GW signal in the LISA range. We  discuss this possibility in Section \ref{sec:LISA}.}  The sensitivity of PTA measurements will soon dramatically improve  
with the SKA experiment.

There are two mechanisms responsible for a SGWB at these scales, associated with PBH. One is the production of GW from the enhanced curvature perturbations when they re-enter the horizon after inflation. This is a completely general mechanism, present for all models of PBH. The second mechanism is production during inflation, from the same inflationary physics that was responsible for the scalar curvature modes. Both  these GW signals are generated at scales corresponding to those of the enhanced scalar modes. Their precise location and amplitude, when compared with the location and amplitude of the PBH mass distribution, can offer experimental information on the specific inflationary mechanism responsible for the PBH generation, and well as on the PBH evolution (via merging and accretion) after they are formed. 

The amplitude of the GW signal is a direct probe of the statistics of the scalar perturbations produced during inflation. In this work we considered two different models that produce the same current PBH distribution (see Figure \ref{fig:f(M)PTA}). One model is characterized by a peak of scalar and primordial GW modes that originate from a rolling axion during inflation, where the scalar curvature modes 
obey a $\chi^2$ statistics. The second model is characterized by Gaussian scalar perturbations, and by the absence of primordial GW. The same PBH distribution at present can be obtained from a much smaller amplitude of scalar perturbations in the Non-Gaussian than in the Gaussian case. As a consequence, the power of the induced GWB (which is proportional to the square of the power of the curvature modes) is much smaller in the Non-Gaussian model. We found that the present PTA sensitivity can be used to discover soon a PBH distribution that significantly contributes to the dark matter of the universe, depending on the precise location of its peak. On the other hand, for the rolling axion model, the induced GW signal is dominated by the primordial one, which is below the current bounds for all values of the peak mass. Most importantly, we found that  both the Gaussian and Non-Gaussian models lead to a signature well above the expected PTA-SKA sensitivity, and therefore could be efficiently used for the detection of this new SGWB from PBH formation.

The location of the SGWB peak, when compared to that of the PBH mass distribution from BHB merging at higher frequencies, will provide important information on the evolution, via merging and accretion, of the PBH mass distribution. While the primordial and induced GW signals probe the perturbations before and during the PBH formation, the current PBH distribution is a function of both the distribution at formation  and the subsequent evolution of the PBH masses. The main components of this evolution are: the mass accretion of each individual PBH from the surrounding plasma, and the primordial black hole merging. We parametrized these effects as a change of the position of the peak mass of the distribution (due to both accretion and merging) as well as an increase of the total mass in PBH (due to accretion). The increase of mass implies that the original PBH distribution was peaked at smaller masses than today. The peak frequency of the primordial and induced GW signal is proportional to the inverse square root of the PBH peak mass at the moment of their formation (see Eq.~(\ref{M-f}) and the subsequent discussion). Therefore, an increase of the peak mass due to accretion and merging implies a smaller original peak mass, and a greater frequency of the primordial and induced GW signal (see the examples shown in Figure \ref{fig:PTAevol} for a measure of this effect, and the corresponding discussion). 

Finally, we also computed the amount of CMB $\mu$-distortion produced by the curvature perturbations responsible for the PBH. In the case of trivial PBH evolution (no merging, nor accretion) the large curvature bump that is required in the Gaussian case produces an amount of distortion below the current  COBE / FIRAS limits, but well above what can be measured in an experiment such as PIXIE. The $\chi^2$ distribution instead produces a distortion which is only marginally greater than that obtained in the absence of a bump, and below the expected PIXIE sensitivity. Accretion and merging imply smaller scale primordial perturbations, thus decreasing the amount of $\mu$-distortion.

To conclude, PBH offer an exciting new possibility for dark matter, with several experi\-mental consequences~\cite{Garcia-Bellido:2017fdg}. They offer a unique window on inflation at scales well below those probed by LSS and CMB observations. The most promising ranges for PBH dark matter are accompanied by an induced (and, depending on the model, primordial) GW signal at either PTA or LISA scales, and well above the sensitivity of these forthcoming experiments. This detection will provide invaluable experimental information on the  specific inflationary mechanism responsible for the PBH generation, and well as on the subsequent PBH evolution.

\vskip.25cm
\section*{Acknowledgements} 

We thank Chiara Caprini, Sebastien Clesse and Kin-Wang Ng for useful discussions. The work of JGB is supported by the Research Project FPA2015-68048-C3-3-P [MINECO-FEDER], and the Centro de Excelencia Severo Ochoa Program SEV-2016-0597.  The work of M.P.  is partially supported from the DOE grant DE-SC0011842  at the University of Minnesota. The work of C.U. is supported by a Doctoral Dissertation Fellowship from the Graduate School of the University of Minnesota.

\vskip.25cm
\section*{Appendices}

\appendix

%%%%%%%%%%%%%%%%%%%%
\section{The $M-N$ relation }
\label{app:MN}
%%%%%%%%%%%%%%%%%%%%

In this Appendix we derive the relation between the PBH mass caused by an overdensity mode produced during inflation, and the number of folds before the end of inflation at which this mode exited the horizon. 

By employing entropy conservation, we can relate the mass $M$ of a PBH to the  wavenumber $k$ of a density mode that collapsed to form the PBH: 
\begin{equation}
M \simeq  20 \, \gamma \,M_{\odot} \left( \frac{k}{10^6 \, {\rm Mpc^{-1}}}  \right)^{-2} \;.  
\label{M-k} 
\end{equation}
See for instance \cite{Nakama:2016gzw} for the derivation of this equation.\footnote{Our numerical value $20$  is about $18\%$ greater than the one of \cite{Nakama:2016gzw} since we are taking $\Omega_{\rm rad} h^2 \simeq 4.2 \times 10^{-5}$, in contrast to the value  $\Omega_{\rm rad} h^2 \simeq 3.6 \times 10^{-5}$ used in \cite{Nakama:2016gzw}. Our numerical value follows from the ratio $\Omega_m h^2 /(1+z_{\rm eq.})$, with the two quantities taken from Table 4 of \cite{Ade:2015xua}. As in  \cite{Nakama:2016gzw}, we take $g_*  = 10.75$ for the number of degrees of freedom in the thermal bath when the mode responsible for the PBH re-enters the horizon. 
In principle, this value does not apply to the full PBH masses that we are considering, however, the value of $M$ in (\ref{M-k}) scales as $g_*^{-1/6}$, and so fixing $g_*  = 10.75$ introduces a negligible error in comparison with the uncertainties associated to PBH collapse.} The density mode collapses into a PBH when the mode re-enters the horizon during the radiation dominated stage. The quantity $\gamma$ is the ratio between the mass collapsing into the PBH and the total mass associated to that mode within the horizon. In our plots we use the numerical value $\gamma = (1/3)^{3/2} \simeq 0.2$ suggested by the analytic computation of \cite{Carr:1975qj}  for a gravitational collapse in the radiation dominated era (see \cite{Carr:2009jm} for a discussion). We stress that the mass value in Eq.~(\ref{M-k}) disregards any PBH mass growth due to  merging or accretion. These effects are discussed in Section \ref{sec:evol}.

We can relate the wavenumber $k$ of the density mode to the number of $e$-folds $N$ before the end of inflation at which this mode exited the horizon. We start by writing $\frac{k}{k_{\rm CMB}} = \frac{a \, H}{a_{\rm CMB} \, H_{\rm CMB}}$, where we take $k_{\rm CMB}=0.002\, {\rm Mpc}^{-1}$ corresponding to the Planck pivot scale. The scale factor during inflation grows as $a = a_{\rm CMB} \,\exp[N_{\rm CMB} - N]$. Using slow roll, we can also write $H = H_{\rm CMB} \,\exp[\epsilon_\phi(N_{\rm CMB} - N)]$, where $\epsilon_\phi = \frac{M_p^2}{2} \left( \frac{1}{V} \, \frac{d V}{d \phi} \right)^2$ is the standard slow-roll parameter.\footnote{Our derivation, and Eq.~(\ref{M-N}), assumes that $\epsilon_\phi$ is constant from $N_{\rm CMB}$ to $N$. The largest value for this interval considered in this work corresponds to modes at LISA scales, for which $N_{\rm CMB} - N \simeq 35$. The parameter $\epsilon_\phi$ changes to second order in slow roll, and, typically,  the quantity $1-\epsilon_{\phi}$ in Eq.~(\ref{M-N}) changes by a negligible amount in this interval. For this reason, we kept $\epsilon_\phi $ constant in our study. It is however straightforward to generalize Eq.~(\ref{M-N}) to the case of varying $\epsilon_\phi$, once a specific inflaton potential is considered.}  Using all this, Eq.~(\ref{M-k}) can be rewritten as 
\begin{equation}
M \simeq 5\times10^{18}  \,M_\odot\,  \gamma \,\, {\rm exp} \left[ 2(N-N_{\rm CMB})(1-\epsilon_{\phi})   \right] \,.
\label{M-N} 
\end{equation}

%%%%%%%%%%%%%%%%%%%%%%%%%%%%%%%%%%%%%
\section{The rolling axion bump model }
\label{sec:model}
%%%%%%%%%%%%%%%%%%%%%%%%%%%%%%%%%%%%%

In this Appendix we summarize the model introduced in \cite{Namba:2015gja} and used in \cite{Garcia-Bellido:2016dkw} to produce PBH from inflation. We refer the interested reader to these works from more details. The lagrangian of the model is  
\begin{equation}
{\cal L} = - \frac{1}{2} \left( \partial \phi \right)^2 - V_{\phi} \left( \phi \right) - \frac{1}{2} \left( \partial \sigma \right)^2  - V_{\sigma} (\sigma) - \frac{1}{4} F^2  - \alpha \frac{\sigma}{4 f_\sigma} F {\tilde F} \;, 
\label{lagrangian-phi-sigma} 
\end{equation}
where $\phi$ is the inflaton field,  $\sigma$ is a pseudo-scalar (axion) spectator field (ie. different from the inflaton and subdominant in energy density) whose motion leads to gauge field amplification, $f_\sigma$ is a mass scale (often denoted as the axion decay constant) and $\alpha$ a dimensionless parameter. The potential for $\sigma$ is chosen to be the simplest one typically associated to  a pseudo-scalar,  
$V_{\sigma} \left( \sigma \right) =  \Lambda^4/2 \left[1+ \cos \frac{\sigma}{f_\sigma} \right] $. The curvature of this potential provides a mass 
$m_\sigma = \Lambda^2/2 f$ in the minimum, and we parametrize by $\delta$ the ratio $\delta \equiv m_\sigma^2/3 H^2$, where $H$ is the Hubble rate during inflation, and we fix $\delta \la 1$. It is easy to verify that, for this choice, the axion has a significant evolution in this potential for a number of $e$-folds equal to $\Delta N \simeq 1/\delta$ \cite{Namba:2015gja}. 

The motion of the axion amplifies gauge field modes that exit the horizon during this time interval. These modes source both scalar and tensor primordial perturbations of comparable wavelength. This results in a bump in the scalar and tensor perturbations, near the modes that left the horizon in this period, and well fitted by the second terms in the relations (\ref{eq:peakedPS}) and (\ref{eq:peakedGW}). The spectra are characterized by three free parameters of the model: (i) the time during inflation at which the roll of $\sigma$ occurs; this is immediately related to the position of the peaks $k_{s,{\rm peak}}$ and $k_{t,{\rm peak}}$ in (comoving) momentum space; (ii) the number of $e$-folds  $\Delta N \simeq 1/\delta$ for which the roll takes place; this is immediately related to the width $\sigma_s^2$ and $\sigma_t^2$ of the two peaks; (iii) the combination $\xi_* \equiv \alpha \, \vert \dot{\sigma}_* \vert/(2 f_\sigma H)$, to which the height of the two peaks is exponentially sensitive; in this relation, $\sigma_*$ is the maximum speed attained by $\sigma$ in its slow-roll. Typically, the PBH limits are saturated for $\xi_* \sim 5-6$. The explicit relations between the model parameters and the power spectra  (\ref{eq:peakedPS}) and (\ref{eq:peakedGW}) are given in Eq.~(4.5) and (4.6) of \cite{Garcia-Bellido:2016dkw}. 

Ref.~\cite{Peloso:2016gqs} studied under which conditions this mechanism is under perturbative control. Only the two cases $\delta = 0.2$ and $\delta = 0.5$ were studied in that work. It was found that perturbativity requires 
\begin{eqnarray}
&&\left( \frac{\Omega_{\rm GW} \, h^2}{2 \times 10^{-9} } \right)_{\rm peak}^{1/4}   \la \frac{f_\sigma}{M_p} \la 1 \;\;,\;\; {\rm for \;} \delta = 0.2 \;, \nonumber\\ 
&&\left( \frac{\Omega_{\rm GW} \, h^2}{6 \times 10^{-8} } \right)_{\rm peak}^{1/4}   \la \frac{f_\sigma}{M_p} \la 1  \;\;,\;\; {\rm for \;} \delta = 0.5 \;,
\label{condition} 
\end{eqnarray} 
where $\Omega_{\rm GW} $ is the fractional energy density in the sourced GW per logarithmic $k$ interval.  We see that the case $\delta = 0.2$ is the more restrictive of the two. In general, we expect the lower bound to be a smooth function of $\delta$.  In this work we consider values of $\delta$ in the $\left[ 0.2 ,\, 0.5 \right]$ interval. All cases we studied satisfy the condition (\ref{condition}) (in fact, they all satisfy the first among  (\ref{condition}), which is the stronger condition in this interval).

%%%%%%%%%%%%%%%%%%%%%%%%%%%
\section{Stochastic GW Backgrounds (SGWB) }
\label{sec:SGWB}
%%%%%%%%%%%%%%%%%%%%%%%%%%%

We denote as $h_{ij}$ the transverse and traceless spatial components of the metric perturbations. These modes, when inside the horizon, satisfy $ k^2 \vert  h_{k} (\eta) \vert^2 = \vert h_{k}^{'}(\eta) \vert^2 $ (where prime denotes derivative wrt conformal time) and their 
energy density per logarithmic wavenumber is 
\begin{eqnarray}
\Omega_{\rm GW} (k, \eta) \! &&= \frac{1}{\rho_{c}}\frac{d \, \rho_{\rm GW}}{d\, \ln k} = \frac{1}{3 H^2 M_p^2} \frac{ M_p^2 }{4 a^2 } \, \frac{d }{d\, \ln k}  \,  \langle \,  \overline{h_{ij}^{'2}}  \, \rangle = \frac{1}{12\, a^2 H^2 } \, \frac{d }{d\, \ln k} \,  \left \langle \,  \overline{ (\partial_{\vec x} h_{ij})^{2}}  \, \right \rangle  \nonumber\\
&&= \frac{1}{12\, a^2 H^2 }  \,\frac{d }{d\, \ln k} \left \langle  \overline{\left( \,  \int  \frac{d^3k  }{(2\pi)^{3/2}} \, \, i \vec k \,\, {\rm e}^{i\, \vec k \cdot \vec x} \,  \sum_\lambda \, \Pi_{ij,\lambda}^* (\vec k )  \,  \hat h_{\lambda}(\vec k, \eta) \right)^2} \,  \right \rangle \,. 
\label{eq:OGWfourier}
\end{eqnarray}
In this expression, $\overline f \left( t \right)$ denotes the time average of the oscillating function $f \left( t \right)$ over one period. 
We have also decomposed the GW into positive and negative helicities $h_\lambda$ (the explicit expression for the  $\Pi_{ij,\lambda}$ projectors can be found for instance in ref. \cite{Namba:2015gja}). 

Once evaluated at the current time $\eta_0$, the expression (\ref{eq:OGWfourier}) gives 
\begin{equation}
\Omega_{\rm GW} (k, \eta_0) \simeq \frac{\Omega_{{\rm rad},0}\, k^2}{24 \,a^2 H^2 }  \sum_\lambda P_{\lambda} \;, 
\end{equation}
where 
\begin{equation}
\langle \hat h_{\lambda}(\vec k, \eta)  \,\,  \hat h_{\lambda'}(\vec k', \eta) \rangle \equiv \frac{2\pi^2}{k^3} \delta^3 (\vec k + \vec k') \, \delta_{\lambda \, \lambda'} \,  P_{\lambda} \,, 
\end{equation}
is the GW power spectrum. 

In models of PBH, various physical processes contribute to the stochastic GW background, as discussed at the beginning of Section 
\ref{sec:GW}.  Throughout this work, we mostly concentrate on the ``primordial'' GW $h_p$ produced during inflation, and on the ``induced'' GW $h_i$ sourced by the scalar primordial perturbations when they re-enter the horizon during the radiation dominated era. 
Accounting also for the (negligible) vacuum GW signal $h_v$ produced during inflation, this results into 
\begin{eqnarray}
\Omega_{\rm GW} (k, \eta_0) &=& \frac{\Omega_{{\rm rad},0}\, k^2}{24 \,a^2 H^2 }  \left( \sum_\lambda P_{h_{v, \lambda}} +  P_{h_{p, \lambda}} +  2 P_{h_p \, h_{i, \lambda}} +  P_{h_{i, \lambda}} \right)    \nonumber\\
&\simeq& \frac{\Omega_{{\rm rad},0}\, k^2}{24 \,a^2 H^2 } \bigg(   P_{h_{p,+}} +  P_{h}^{pi}   +  \sum_\lambda P_{h_{v, \lambda}} + P_{h_{i, \lambda}}  \bigg) \,. 
\label{Omega-GW}
\end{eqnarray}

We note that the vacuum and the induced GW signals are helicity-independent. On the contrary, one helicity of the primordial GW is much smaller than the other one, and can be disregarded. The cross-correlation $ P_{h}^{pi} $ includes only the dominant helicity.

\section{Auto-Correlation $\left\langle h_i h_i \right\rangle$ and the zero-width approximation }
\label{appsec:induced}

In this section we compute the diagrams shown in Figure~\ref{fig:hihi}, contributing to the power spectrum~(\ref{indgwNG}) of the 
induced GW. In the first of the three diagrams we note the presence of two internal power spectra of the sourced scalar perturbations, given by the second term in (\ref{eq:peakedPS}). Using this result, the diagram becomes a one loop diagram, which we evaluate numerically through Eq.~(\ref{indgwG}). The other two diagrams are genuinely  3-loop diagrams, so in addition to the two time integrals shown explicitly in Eq.~(\ref{indgwNG}), they involve  $8$ dimensional integrals in momentum space.\footnote{We start from $3$ arbitrary internal momenta, parametrized by $3$ coordinates each; without loss of generality, we are free to orient the external momentum ${\bf k}$ along the $z-$axis, and to orient one of the internal momenta, say ${\bf p}_1$ in the $x-z$ plane, multiplying the final result by $2 \pi$ (by statistical isotropy, this is equivalent to performing one integration in which all the internal momenta rotate, in such a way that the internal momentum ${\bf p}_1$ rotates at a fixed angle with respect to ${\bf k}$).} 

Rather than performing a $10$ dimensional integration, we estimate the last two diagrams by exploiting the fact that the sourcing gauge fields are very peaked in momentum space. The exact amplitude of the vector fields in the rolling axion bump model is given by~\cite{Namba:2015gja,Peloso:2016gqs}
\begin{equation}
A_+ \left( \tau, k \right) \simeq  N \left[ \xi_* ,\, x_* \equiv \frac{k}{k_*} ,\, \delta \right] \, \left( \frac{-\tau}{8 \, k \, \xi \left( \tau \right)} \right)^{1/4}  \, {\rm exp} \left[ - \frac{4 \xi_*^{1/2}}{1+\delta} \, \left( \frac{-\tau}{-\tau_*} \right)^{\delta/2} \, \left( - k \tau \right)^{1/2} \right] \;, 
\label{wfgauge}
\end{equation}
where $\tau$ is conformal time (which is negative during inflation, $\tau \simeq -1/{\cal H}$) and $\tau_*$ is the time at which the axion rolls fastest, and we have used $\xi(\tau) = \alpha \,|\sigma'|/(2 f {\cal H})$, with $\xi_* \equiv \xi \left( \tau_* \right)$. 
The normalization factor can be well fitted by a log-normal shape as shown in \cite{Peloso:2016gqs} 
\begin{equation}
N \left[ \xi_* ,\, x_* \equiv \frac{k}{k_*} ,\, \delta \right] \approx N_c \left[ \xi_* ,\, \delta \right] \, {\rm exp} \left[ - \frac{1}{2 \, \sigma_A^2 \left[ \xi_* ,\, \delta \right] } \, \ln^2 \left( \frac{x_*}{q_A^c \left[ \xi_* ,\, \delta \right] } \right) \right] \;,  
\label{Nfit}
\end{equation}
and the functional dependence of $N_c ,\, \sigma_A^2 ,\, q_A^c$ on $\xi_*$ and $\delta$ is given in  Ref.~\cite{Peloso:2016gqs}. The parameter $q_A^c$ is an order-one number, and therefore the gauge field amplitude (\ref{Nfit}) exhibits a peak in momentum space at $k \sim k_*$, namely at the scales that left the horizon when the axion rolled fastest. As the gauge fields source both scalar and tensor modes, this peak  is the origin of the bumps in the scalar perturbations and gravitational waves produced in this model. 

In the zero-width approximation, we replace the exact gauge field profile with a Dirac delta function, with support at $k = q_A^c \, k_*$. 
More concretely, the approximation is done after writing the gauge field propagators. Each propagator is proportional to $N^2$, and we approximate 
\begin{equation}
N^2 \left( \frac{p}{k_*}, \xi_*,\delta \right) \;\; \rightarrow \;\; N^2_{\rm app} \left( \frac{p}{k_*} , \, \xi_*, \, \delta \right) \equiv {\cal N}^2 \left( \xi_* ,\, \delta \right) \, \delta \left( \ln \frac{p}{ q_A^c \, k_* } \right) \,, 
\end{equation} 
where $p$ is the magnitude of the internal momentum of the corresponding gauge field, and where $ {\cal N}^2 $ must be chosen so that the integral over $p$ using the approximate propagator gives a good estimate of the exact integral. 

The amplitudes can be written as the appropriate power of the external momentum time dimensionless ratios involving external and internal momenta. As all the gauge fields wavefunctions are peaked at $p =   q_A^c \, k_*$, all the internal momenta in the diagrams are comparable to each other, and comparable to the external momentum. Therefore each internal momentum integral has a measure of the type $\int d p/p$ times an order-one factor. We use this measure to determine ${\cal N}$. Specifically, we require that 
\begin{equation}
\int \frac{dp}{p}  \; N^2 \left(\frac{p}{k_*}, \, \xi_*, \, \delta \right) = \int \frac{dp}{p} \; N_{\rm app}^2 \left( \frac{p}{k_*} , \, \xi_* , \, \delta \right)  \;. 
\end{equation}
Using the expression (\ref{Nfit}), the integration at the left hand side gives $\sqrt{\pi} N_c^2 \sigma_A$. The integral at the right hand side is by definition ${\cal N}^2$. Therefore, our zero-width approximation consists in writing the exact amplitude, and then replace 
\begin{equation} 
N^2 \left( \frac{p}{k_*}, \xi_*,\delta \right) \;\; \rightarrow \;\; \pi^{1/2} \, N_c^2 \left( \xi_*,\delta \right) \;  \sigma_A \left[ \xi_* ,\, \delta \right]  \, \delta \left( \ln \frac{p}{ q_A^c \, k_* } \right) \,,  
\label{0width} 
\end{equation} 
in each gauge field propagator. As there are four gauge field propagators in each diagram, this introduces four $\delta$-functions, that considerably simplify the 8-dimensional momentum integration. 

The prescription (\ref{0width})  is just one simplifying procedure, among many possible ones. Eventually, its goodness can only be understood by comparing exact and approximate results. We can do so for the Reducible diagram (the first in Figure  \ref{fig:hihi}), for which we have the exact result (obtained by evaluating the one loop expression (\ref{indgwG})) as well as the approximate one (given in the next subsection). The exact and the approximate results for this diagram are shown in Figure \ref{fig:primvsinduced}) where one can see that they lead to two bumps in the GW signal that differ from each other by a factor of $\sim 2$ (both in the amplitude and the position of the peak). This allows us to conclude that the zero-width approximation (\ref{0width}) can indeed be used to estimate the order of magnitude of the second and third diagram of Figure \ref{fig:hihi}.

The remainder of this appendix is divided into three parts, in which we give the explicit expression of the integrand of the three diagrams of Figure~\ref{fig:hihi} in the zero-width approximation.

%%%%%%%%%%%%%%%%%%%%%%%%%%%%%%%%%
\subsection{Reducible diagram} 
%%%%%%%%%%%%%%%%%%%%%%%%%%%%%%%%%

We now give the explicit expression of the integrand of the Reducible diagram in the  zero-width approximation. Referring to the first diagram in Figure  \ref{fig:hihi}, we label the external momentum as ${\bf k}$ (going from left to right in the diagram), and we label the internal momenta in the first, second, and third gauge field propagator (from top down), as ${\bf u}$, ${\bf n}$, and ${\bf v}$, respectively (all going from left to right in the diagram). All the other momenta in the diagram can be obtained from these ones, using momentum conservation at each vertex. With this parametrization we can immediately perform the three integrals $\int du \int dn \int dv$ using the Dirac $\delta$-function in (\ref{0width}). 

Using this diagram in Eq.~(\ref{indgwNG}), and rescaling the final result as in Eq.~ (\ref{Omega-GW}), we obtain 
\begin{eqnarray}
&& \Omega_{\rm GW}^{\rm Reducible}h^2 (t_0)  =  \frac{\Omega_{{\rm rad},0} \; h^2(k \eta_{\rm eval})^2}{12} \;  \frac{(PF_{\zeta})^4 \; (N_c(\xi_* ,\, \delta) \; \sqrt{\pi} \; \sigma_A)^4}{(2  \pi)^{11} } \times \nonumber\\  
&&\int d \Omega_u \, d \Omega_n  \, d \Omega_v \, 
 \left( 1 - \left(  \frac{{\bf k}\cdot ( {\bf u}+ {\bf n} ) }{ k\; \vert  {\bf u+n}  \vert} \right)^2   \right)^2  \left(  \frac{4 {p_c}^2 - \vert  {\bf u + n} \vert^2 }{2 {p_c}^2 }    \right)^2 \;\; 
\left(  \frac{4 { p_c}^2 - \vert   {\bf k - u -n }  \vert^2 }{2 {p_c}^2 }  \right)^2 \times
\nonumber\\  
&& 
\hspace{1cm}\frac{p_c^{14} \,\delta \left( \vert  {\bf k  - u - n - v}  \vert  -  p_c  \right)}
{k \;  \vert {\bf u+n}  \vert^4 \;\;  \vert {\bf k-u-n} \vert^8}  \ \ 
{\cal T}_{\rm ind}^{ \, 2} \left(  \frac{ \vert   { \bf u+n } \vert}{k}, \frac{ \vert  {\bf k-u-n} \vert }{k} , \,\, \eta_{\rm eval} \right)  \times \nonumber\\
&& \hspace{1cm}
{\cal T}_{\zeta}^2  \left( \vert {\bf u + n}  \vert, \; 2\sqrt{\frac{ p_c}{ \vert  {\bf u + n} \vert }} \right)  \;\; 
{\cal T}_{\zeta}^2  \left ( \vert {\bf k-u-n}  \vert   , \; 2\sqrt{\frac{ p_c}{ \vert {\bf k-u-n}   \vert }} \right) \;,  
\label{Reducible}
\end{eqnarray}
where $p_c = q_A^c k_*$ (the support of the $\delta$-function). We now discuss the various quantities entering in this expression. 

The quantity $PF_{\zeta}$ is a prefactor associated to each scalar propagator, and it reads 
\begin{equation} 
PF_{\zeta} \equiv \frac{3 \, \pi^{3/2} \; H^2 \; \alpha \; \delta}{8 \, \sqrt{2} \; M_p^2}   \,. 
\end{equation} 
In Eq.~(\ref{Reducible}), it multiplies the prefactor of each of the four $\delta$-functions arising from the zero-width approximation. We have used three of these $\delta$-function to reduce the integral over internal momenta into a six-dimensional angular integral. We note that one $\delta$-function (the one associated with the bottom gauge field propagator in the diagram) has not yet been imposed, and it is still present in Eq.~(\ref{Reducible}). We also note that one angular integral can be trivially done, as discussed at the beginning of this Appendix. 

The first parenthesis on the second line of Eq.~(\ref{Reducible}) arises from the polarization operators of the gravitational waves; the next two parenthesis arise from the polarization operators of the gauge fields in the propagators. The product ${\cal T}_{\rm ind}^2$ accounts for the two time integrations in the beginning Eq.~(\ref{indgwNG}); each term is given by 
\begin{equation}
{\cal T}_{\rm ind} (  A,  B,  \eta_{\rm eval} ) \equiv \frac{16}{9 \,\, (x_{\rm eval})} \int_0^{x_{\rm eval}} dx'  x' \, \sin (x_{\rm eval}-x') \, F_T  \left( A \, x',  B \, x'   \right)
\end{equation}
where $F_T$ is given in Eq.~(\ref{F-T}), and where $ x_{\rm eval} \equiv k \eta_{\rm eval} $, with $\eta$ being conformal time. For each mode, we choose $ x_{\rm eval} = 20$, namely we evaluate this expression when the GW mode is well within the horizon (at this moment, the $\zeta \zeta \rightarrow h_i$ has at all effects concluded, namely $h$ evolves as if it was a free mode). 

Finally, the two quantities ${\cal T}_{\zeta}^2 $ are time integrals arising from the four propagators of the scalar perturbations, and account for the fact that these modes are sourced by the gauge fields during inflation.  Their explicit expressions is given in Eq.~(C.8) of   \cite{Namba:2015gja}.

%%%%%%%%%%%%%%%%%%%%%%%%%%%%%%%%%%
\subsection{Planar Diagram} 
%%%%%%%%%%%%%%%%%%%%%%%%%%%%%%%%%

We now give the explicit expression of the integrand of the Planar diagram in the  zero-width approximation. Referring to the second  diagram in Figure  \ref{fig:hihi}, we label the external momentum as ${\bf k}$ (going from left to right in the diagram), the internal momentum of the top gauge field propagator as ${\bf u}$  (going from left to right), the internal momentum of the left  gauge field propagator as ${\bf n}$  (going from up to down) and the internal momentum of the right gauge field propagator as ${\bf z}$ (going from down to up). Proceeding as in the previous subsection, we obtain 
\begin{eqnarray}
&& \!\!\!\!\!\!\!\!\!\!\!\!\!\!\!\!\!\!  \Omega_{\rm GW}^{\rm PD}h^2 (t_0)  =  \frac{\Omega_{{\rm rad},0} \; h^2(k \eta_{\rm eval})^2}{12}  \times 32 \; \frac{(PF_{\zeta})^4 \; (N_c(\xi_* ,\, \delta) \; \sqrt{\pi} \; \sigma_A)^4}{(2  \pi)^{11} } \int d\Omega_u \, d \Omega_n  \, d \Omega_z \,  
\left( 1 - \left(  \frac{{\bf k}\cdot ({\bf u+n}) }{ k\; \vert {\bf u+n}  \vert} \right)^2   \right) 
 \nonumber\\  \nonumber\\  
&& 
 \left( 1 - \left(  \frac{{\bf k}\cdot ({\bf u + z}) }{ k\; \vert  {\bf u + z}  \vert} \right)^2   \right) \;\;\; 
 \frac{ {\cal T}_{\rm ind} \left( \frac{ \vert  {\bf u+n}  \vert } {k}, \frac{ \vert   {\bf k - u - n}   \vert} {k} , \,\, \eta_{\rm eval} \right)   \;\; 
{\cal T}_{\rm ind} \left( \frac{ \vert {\bf u + z} \vert } {k},  \frac{ \vert  {\bf k-u-z}  \vert }{k} , \,\, \eta_{\rm eval} \right)  \;\; { p_c}^{14} }{{ k}^2 \;  \vert {\bf u+n} \vert^2 \;\;  \vert   {\bf k-u-n}  \vert^4   \;  \vert   {\bf u+z}  \vert^2 \;\;  \vert   {\bf k-u-z}  \vert^4 } \;\;
 \nonumber\\  \nonumber\\  
&& 
{\cal T}_{\zeta}  \left( \vert {\bf u+n}   \vert, \; 2\sqrt{\frac{ p_c}{ \vert {\bf u+n}  \vert }} \right)  \;\; 
{\cal T}_{\zeta}  \left(   \vert   {\bf k-u-n}   \vert   , \; 2\sqrt{\frac{ p_c}{  \vert  {\bf k-u-n}   \vert }} \right) 
 {\cal T}_{\zeta}  \left( \vert  {\bf u+z}  \vert, \; 2\sqrt{\frac{ p_c}{ \vert  {\bf u+z} \vert }} \right)  \;\; 
 \nonumber\\  \nonumber\\  
 &&  
{\cal T}_{\zeta}  \left(   \vert   {\bf k-u-z} \vert   , \; 2\sqrt{\frac{ p_c}{  \vert {\bf k-u-z}  \vert }} \right) \;\;
\delta \left( \cos \theta_{uk} - \frac{{ k}}{2{ p_c}}   \right) \times \epsilon_{\rm planar} \;, 
\end{eqnarray}
where 
\begin{equation}
\epsilon_{\rm planar}= \left( {\vec \epsilon}^+ ({\bf u}) \cdot {\vec \epsilon}^+ ({\bf n})\right) \;\;
 \left( {\vec \epsilon}^+ ({\bf -n}) \cdot {\vec \epsilon}^+ ({\bf k-u})\right) \;\; 
\left( {\vec \epsilon}^+ ({\bf -u}) \cdot {\vec \epsilon}^+ ({\bf -z})\right) \;\;
 \left( {\vec \epsilon}^+ ({\bf z}) \cdot {\vec \epsilon}^+ ({\bf u-k})\right) \;, 
\end{equation}
arising from polarization operators of the vector fields, and where the other quantities have been given in the 
previous subsection.

%%%%%%%%%%%%%%%%%%%%%%%%%%%%%%%%%
\subsection{Non-Planar Diagram} 
%%%%%%%%%%%%%%%%%%%%%%%%%%%%%%%%%

We now give the explicit expression of the integrand of the Non-Planar diagram in the  zero-width approximation. Referring to the third diagram in Figure  \ref{fig:hihi}, we label the external momentum as ${\bf k}$ (going from left to right in the diagram), the internal momentum of the gauge field propagator going from top left to top right as ${\bf u}$,  the internal momentum of the gauge field propagator going from top left to bottom right as ${\bf n}$,  and the  internal momentum of the gauge field propagator going from bottom left to top right as ${\bf v}$.   Proceeding as in the previous subsections, we obtain 
\begin{eqnarray}
&& \!\!\!\!\!\!\!\!\!\!\!\!\!\!\!\!\!\!   \Omega_{\rm GW}^{\rm NPD}h^2 (t_0)  =  \frac{\Omega_{{\rm rad},0} \; h^2(k \eta_{\rm eval})^2}{12}  \times 32 \; \frac{(PF_{\zeta})^4 \; (N_c(\xi_*) \; \sqrt{\pi} \; \sigma_A)^4}{(2  \pi)^{11} } \int d \Omega_u \, d \Omega_n  \, d \Omega_v \,   
\left( 1 - \left(  \frac{{\bf k}\cdot ( {\bf u + n} ) }{ k\; \vert  {\bf u + n} \vert} \right)^2   \right) 
\nonumber\\ \nonumber\\  
&& 
 \left( 1 - \left(  \frac{{\bf k}\cdot (  {\bf u + v} ) }{ k\; \vert   {\bf u + v}  \vert} \right)^2   \right)   \;\;\;
 \frac{ {\cal T}_{\rm ind} \left( \frac{ \vert   {\bf u + n} \vert}  {k} , \frac{ \vert  {\bf k-u-n}  \vert } {k} , \,\, \eta_{\rm eval} \right)   \;\;
{\cal T}_{\rm ind} \left( \frac{ \vert   {\bf u + v} \vert}  {k} ,  \frac{ \vert  {\bf k-u-v} \vert  } {k} , \,\, \eta_{\rm eval} \right)  \;\; { p_c}^{14} } {{ k} \;  \vert {\bf u + n} \vert^2 \;\;  \vert   {\bf k-u-n} \vert^4   \;  \vert  {\bf u + v} \vert^2 \;\;  \vert  {\bf k-u-v}  \vert^4 } 
 \nonumber\\  \nonumber\\  
&&
{\cal T}_{\zeta}  \left( \vert  {\bf u + n} \vert, \; 2\sqrt{\frac{ p_c}{ \vert  {\bf u + n}  \vert }} \right)  \;\; 
{\cal T}_{\zeta}  \left(   \vert   {\bf k-u-n}  \vert   , \; 2\sqrt{\frac{ p_c}{  \vert  {\bf k-u-n} \vert }} \right) \;\;
{\cal T}_{\zeta}  \left( \vert {\bf u + v}  \vert, \; 2\sqrt{\frac{ p_c}{ \vert {\bf u + v}  \vert }} \right)  \;\; 
\nonumber\\   \nonumber\\  
&&
{\cal T}_{\zeta}  \left(   \vert  {\bf k-u-v}  \vert   , \; 2\sqrt{\frac{ p_c}{  \vert {\bf k-u-v}  \vert }} \right) \;\;
\delta \left( \vert  {\bf k-u-n-v}   \vert -  p_c  \right) \;\; \epsilon_{\rm nonplanar}  \;, 
\end{eqnarray}
with
\begin{eqnarray}
 \epsilon_{\rm nonplanar} &&= \left( {\vec \epsilon}^+ ({\bf u}) \cdot {\vec \epsilon}^+ ({\bf n})\right) \;\;
 \left( {\vec \epsilon}^+ ({\bf v}) \cdot {\vec \epsilon}^+ ({\bf k-u-n-v})\right) \times
 \nonumber \\[2mm]
&&\hspace{5mm}
\left( {\vec \epsilon}^+ ({\bf -u}) \cdot {\vec \epsilon}^+ ({\bf -v})\right) \;\;
 \left( {\vec \epsilon}^+ ({\bf -n}) \cdot {\vec \epsilon}^+ ({\bf -(k-u-n-v)})\right) \;, 
\end{eqnarray}
again arising  from polarization operators of the vector fields.

\section{Cross Correlation $\left\langle h_p h_i \right\rangle$ }
\label{appsec:cross}

In this appendix we provide the expression for the mixed correlator between the primordial and the induced GW. 
We start from Eq.~(\ref{eq:primindgw}) and rewrite the result in terms of its contribution to Eq.~ (\ref{Omega-GW}). The contribution is expressed by the second and third diagram of Figure  \ref{fig:hphp-hphi}. Referring to the third diagram, we label the external momentum as ${\bf k}$ (going from left to right in the diagram), the internal momentum of the upper scalar propagator as ${\bf p}$ (going from left to right), and the internal momentum of the upper gauge field propagator as ${\bf q}$ (going from right to left). The two diagrams give an identical result, and their sum is given by the integral 
\begin{eqnarray}
&& \!\!\!\!\!\!\!\!\!\!\!\!\!\!\!\!\!\! \!\!\!\!\!\!\!\! 
 \Omega_{\rm GW}^{\rm mixed-term}h^2 (t_0)  =  \frac{\Omega_{{\rm rad},0} \; h^2(k \eta_{\rm eval})^2}{24}  \times 16 \; \frac{ (PF_{h}) \;(PF_{\zeta})^2 }{(2  \pi)^{8} (k \eta_{\rm eval})} \int d^3 { p} \; d^3 { q} \; N^2 \left( \frac{q}{k_*},\, \xi_*,\delta \right) \; N^2 \left( \frac{\vert {\bf p+q} \vert}{k_*}, \xi_*,\delta \right)\; \nonumber\\   \nonumber\\  
&& \!\!\!\!\!\!\!\!\!\!\!\!\!\!\!\!\! \!\!\!\!\!\!\!\! 
N^2 \left( \frac{\vert {\bf k+q } \vert} {k_*}, \xi_*,\,  \delta \right) \;
\frac{ \vert {\bf p+q}  \; \vert^{1/2} { q}^{1/2}  \;  \vert {\bf k+q} \vert^{1/2}  ({ q}^{1/2} + \vert {\bf p+q} \vert^{1/2}  ) \; (\vert {\bf k+q} \vert^{1/2}   +  \vert {\bf p+q } \vert^{1/2}  )  }{ k^{1/2} \; p^4  \; \vert  {\bf k-p }  \vert^4}    \;\;
{\cal T}_{\rm ind} \, \left(   \frac{p}{k} , \,  \frac{\vert {\bf k-p} \vert}{k} \, , \, \eta_{\rm eval} \right) \;\;\;
 \nonumber\\  \nonumber\\  
&& \!\!\!\!\!\!\!\!\!\!\!\!\!\!\!\!\!\!\ \!\!\!\!\!\!\!\! 
 {\cal T}_{h}  \left(  \frac{k}{k_*} ,  \frac{\sqrt{ q} + \sqrt{\vert {\bf k+q} \vert}} {\sqrt{k} } \right)  \;
{\cal T}_{\zeta}  \left( \frac{p}{k_*}, \frac{\sqrt{ q} + \sqrt{\vert  {\bf p+q}   \vert}}{\sqrt{ p} } \right)  \; 
{\cal T}_{\zeta}  \left( \frac{\vert  {\bf  k-p}  \vert} {k_*}, \frac{\sqrt{ \vert {\bf p+q }\vert} + \sqrt{\vert  {\bf k+q}   \vert}}{\sqrt{ \vert  {\bf k-p}  \vert} } \right)  \; \epsilon_{\rm mixed-term}  \;, 
\end{eqnarray}
with
\begin{eqnarray}
\epsilon_{\rm mixed-term} &=& \left( {\vec \epsilon}^+ ({\bf -k}) \cdot  {\bf \hat p}  \right)^2 \;\;
 \left( {\vec \epsilon}^+ ({\bf -q}) \cdot {\vec \epsilon}^+ ({\bf p+q})\right) \;\; 
\left( {\vec \epsilon}^+ ({\bf -p-q}) \cdot {\vec \epsilon}^+ ({\bf k+q})\right) \times
\nonumber\\[2mm]
&&\left( {\vec \epsilon}^+ ({\bf k}) \cdot {\vec \epsilon}^+ ({\bf q})\right) \;\;
\left( {\vec \epsilon}^+ ({\bf k}) \cdot {\vec \epsilon}^+ ({\bf -k-q})\right) \;, 
\end{eqnarray}
arising from the polarization operators of the GW and of the gauge fields. We have also introduced the quantity 
\begin{equation}
PF_h=\frac{\sqrt{2} \; H^2 \; }{ Mp^2} \;, 
\end{equation}
arising from a prefactor in the expression of $h_p$ (see Eq.~(D.6) of  \cite{Namba:2015gja}).

%%%%%%%%%%%%%%%%%%%%
\section{Simple parametrization of the PBH evolution and present abundance  }
\label{appsec:merging}
%%%%%%%%%%%%%%%%%%%%

In this Appendix we derive the relation (\ref{f-A-M}) of the main text. We follow the computations of \cite{Carr:2009jm} up to Eq.~
(\ref{f-noA-noM}), and we then introduce two parameters to account for PBH nontrivial evolution (accretion and merging after their formation). We start by relating time and temperature during radiation domination. In a radiation dominated universe the Hubble rate $H$ is related to time by $H=\frac{1}{2 t}$. From the Friedmann relation $H^2 = \frac{\rho}{3 M_p^2}$, and from the energy density in a thermal bath of temperature $T$, given by $\rho = \frac{\pi^2}{30} \, g_* T^4$, where $g_*$ is the number of effective bosonic degrees of freedom at the temperature $T$, one obtains 
\begin{equation}
t = 0.74 \, {\rm s} \left( \frac{g_*}{10.75} \right)^{-1/2} \, \left( \frac{T}{\rm MeV} \right)^{-2} \;. 
\end{equation} 

We denote by $M_f$ the PBH mass at formation, by $t_f$ the formation time, and by $T_f$ the temperature at this moment. The PBH formation occurs when a mode re-enters the horizon during the radiation dominated era, and the PBH mass at this moment is equal to the mass inside the horizon 
(namely, in the volume $\frac{4 \pi}{3} H^{-3}$)  times an efficiency factor $\gamma = {\cal O } \left( 10^{-1} \right)$. This gives 
\begin{equation}
M_f = 2.0 \times 10^5 \, \gamma \, \left( \frac{t_f}{\rm s} \right) M_\odot = 1.5 \times 10^5 \, \gamma \, M_\odot \, \left( \frac{g_*}{10.75} \right)^{-1/2} \, \left( \frac{T_f}{\rm MeV} \right)^{-2} \;. 
\label{Mf-Tf} 
\end{equation}

For a distribution of PBH formation masses $M_f$ the quantity $\beta$ gives the ratio~\footnote{For notational convenience, we express the functional dependence of $\beta$ on the PBH mass in terms of the ratio $M_f / M_\odot$.}  at the time $t_f$ between the PBH energy density of the mass $M_f$ (more specifically, of the corresponding logarithmic interval) and the background energy density 
\begin{equation}
\beta \left( \frac{M_f}{M_\odot} \right) = \frac{1}{\rho_\gamma} \, \frac{d \rho_{\rm PBH} \left( M_f ,\, t_f \right)}{d \ln M_f} 
 = \frac{4/3}{s \left( t_f \right) T_f} \, \frac{d \rho_{\rm PBH} \left( M_f ,\, t_f \right)}{d \ln M_f} \;, 
\label{beta-rhof}
\end{equation} 
where  $s$ the entropy density of the thermal bath. 

We solve Eq.~(\ref{Mf-Tf}) as an expression for the formation temperature, and we insert it in Eq.~(\ref{beta-rhof}) to write 
\begin{equation}
\frac{d \rho_{\rm PBH} \left( M_f ,\, t_f \right)}{d \ln M_f} = 290 \, \gamma^{1/2} 
\, \sqrt{ \frac{M_{\odot}}{{M}_f}} \; \beta \left( \frac{M_f}{M_\odot} \right) \, s \left( t_f \right) \, {\rm MeV}  \,, 
\label{rho-s}
\end{equation} 
where here and in the following we  fix $g_* = 10.75$, ignoring the weak dependence of $g_*^{1/4}$ on temperature (doing so induces an error that is much smaller than the uncertainties on the other parameters, as for instance $\gamma$ and the merging and accretion parameters introduced below). 

Without PBH accretion and merging, the PBH energy density between the time $t_f$ and the present time $t_0$ scales as the inverse volume of the universe, analogously to the entropy density. Therefore, under this assumption \cite{Carr:2009jm}, Eq.~(\ref{rho-s}) also holds true if we evaluate both $\rho_{\rm PBH}$ and  $s$ at the present time. Doing so, and  dividing the resulting expression by the present energy density of dark matter gives 
\begin{eqnarray} 
f_{\rm PBH} \left( M_f ,\, t_0 \right) & \equiv & \frac{1}{\rho_{\rm CDM} \left( t_0 \right)} \, \frac{d \rho_{\rm PBH} \left( M_f ,\, t_0 \right)}{d \ln M_f} = 
 290 \, \gamma^{1/2} 
\, \sqrt{ \frac{M_{\odot}}{{M}_f}} \; \beta \left( \frac{M_f}{M_\odot} \right) \, \frac{s \left( t_0 \right) \, {\rm MeV}}{ \rho_{\rm crit,0} \, \Omega_{\rm CDM}} \nonumber\\ 
& = & 6.7 \times 10^8 \,  \gamma^{1/2}  \, \sqrt{ \frac{M_{\odot}}{{M}_f}} \; \beta \left( \frac{M_f}{M_\odot} \right) \;\;\;, \;\; {\rm no \; accretion \; nor \; merging} \;, 
\label{f-noA-noM}
\end{eqnarray} 
where $\Omega_{\rm CDM} h^2 = 0.12$ \cite{Ade:2015xua} has been used. We note that in this relation $M_f$ gives both the PBH mass at formation, and today, since accretion and merging are neglected here. 

We extend this relation with a simple parametrization to account for merging and accretion. We assume that each PBH mass increases by a factor ${\cal A} > 1$ due to accretion from the surrounding plasma. Realistically, we expect ${\cal A}$ to be function of the PBH mass and of time. However, a single parameter ${\cal A}$ is a reasonable simplifying assumption if $\beta$ is narrowly peaked. (We also expect that ${\cal A}$ depends on the specific environment where the PBH is located. This is not captured by our parametrization.) Under this assumption, accretion leads to the present value $M = {\cal A} \, M_f$ for the PBH mass, and to the increase of the total energy density in PBH $\rho_{\rm PBH} \left( t_0 \right) = {\cal A} \times \rho_{\rm PBH} \left( t_0 ,\, {\cal A} = 1 \right) $. We then parametrize merging also with a single parameter, by assuming that merging shifts the peak of PBH distribution by a factor ${\cal M} > 1$, without changing its shape, nor the total energy density in PBH. 

As a consequence of these effects, the distribution (\ref{f-noA-noM}) is changed into 
\begin{equation} 
f_{\rm PBH} \left( M ,\, t_0 ,\, {\cal A} ,\, {\cal M} \right) \simeq  6.7 \times 10^8 \,  \gamma^{1/2}  {\cal A} \, \sqrt{ \frac{{\cal A} \, {\cal M} \, M_{\odot}}{{M}}} \; \beta \left( \frac{M}{{\cal A} \, {\cal M}\,M_\odot } \right) \;. 
\label{f-A-M-app}
\end{equation} 
This expression reduces to (\ref{f-noA-noM}) in the case of no accretion and merging, ${\cal A} = {\cal M} = 1$. 

In the single mass limit, $\beta \left( x \right) \propto \delta \left( x - {\bar m} \right)$, (where $\delta$ is the Dirac delta function) the shift changes the PBH mass from its formation value $M_f = {\bar m} \, M_\odot$ to the present value $M =  {\cal A} \, {\cal M} \, {\bar m} \,M_\odot$. In generality (namely, for an arbitrary distribution), the full distribution $\beta$ is shifted towards greater values without changing shape. Finally, the total energy associated with (\ref{f-A-M-app}) is 
\begin{eqnarray} 
&& \!\!\!\!\!\!\!\!  \!\!\!\!\!\!\!\! 
\frac{\rho_{\rm PBH} \left( t_0 ,\, {\cal A} ,\, {\cal M} \right)}{\rho_{\rm CDM} \left( t_0 \right)}  =   \int \frac{dM}{M} \; f_{\rm PBH} \left( M ,\, t_0 ,\, {\cal A} ,\, {\cal M} \right) \nonumber\\ 
&&  =  {\cal A} \times \int \frac{dM}{M} \; f_{\rm PBH} \left( M ,\, t_0 ,\, {\cal A} = 1 ,\, {\cal M} = 1 \right) = {\cal A} \times \frac{\rho_{\rm PBH} \left( t_0 ,\, {\cal A} = 1 ,\, {\cal M} = 1 \right)}{\rho_{\rm CDM} \left( t_0 \right)} \;,  
\label{F-A}
\end{eqnarray} 
namely it is increased by accretion, but not by merging, as required.


\begin{thebibliography}{99}



\bibitem{Garcia-Bellido:2017fdg} 
  J.~Garc\'ia-Bellido,
  %``Massive Primordial Black Holes as Dark Matter and their detection with Gravitational Waves,''
  J.\ Phys.\ Conf.\ Ser.\  {\bf 840}, no. 1, 012032 (2017)
  %doi:10.1088/1742-6596/840/1/012032
  [arXiv:1702.08275 [astro-ph.CO]].
  %%CITATION = doi:10.1088/1742-6596/840/1/012032;%%
  %10 citations counted in INSPIRE as of 09 Jun 2017
  

 
%\cite{GarciaBellido:1996qt}
\bibitem{GarciaBellido:1996qt} 
  J.~Garc\'ia-Bellido, A.~D.~Linde and D.~Wands,
  %``Density perturbations and black hole formation in hybrid inflation,''
  Phys.\ Rev.\ D {\bf 54}, 6040 (1996)
  %doi:10.1103/PhysRevD.54.6040
  [astro-ph/9605094].
  %%CITATION = doi:10.1103/PhysRevD.54.6040;%%
  %230 citations counted in INSPIRE as of 06 Oct 2016


 %\cite{Clesse:2015wea}
\bibitem{Clesse:2015wea} 
  S.~Clesse and J.~Garc\'ia-Bellido,
  %``Massive Primordial Black Holes from Hybrid Inflation as Dark Matter and the seeds of Galaxies,''
  Phys.\ Rev.\ D {\bf 92}, no. 2, 023524 (2015)
  %doi:10.1103/PhysRevD.92.023524
  [arXiv:1501.07565 [astro-ph.CO]].
  %%CITATION = doi:10.1103/PhysRevD.92.023524;%%
  %16 citations counted in INSPIRE as of 29 Sep 2016
  
 
  %\cite{Clesse:2016vqa}
\bibitem{Clesse:2016vqa} 
  S.~Clesse and J.~Garc\'ia-Bellido,
  %``The clustering of massive Primordial Black Holes as Dark Matter: measuring their mass distribution with Advanced LIGO,''
  Phys.\ Dark Univ.\  {\bf 15}, 142 (2017)
  %doi:10.1016/j.dark.2016.10.002
  [arXiv:1603.05234 [astro-ph.CO]].
  %%CITATION = doi:10.1016/j.dark.2016.10.002;%%
  %55 citations counted in INSPIRE as of 26 Jun 2017


 
  
    %\cite{Ezquiaga:2017fvi}
\bibitem{Ezquiaga:2017fvi} 
  J.~M.~Ezquiaga, J.~Garc\'ia-Bellido and E.~Ruiz Morales,
  %``Primordial Black Hole production in Critical Higgs Inflation,''
  arXiv:1705.04861 [astro-ph.CO].
  %%CITATION = ARXIV:1705.04861;%%
  %1 citations counted in INSPIRE as of 22 May 2017
  
  
  %\cite{Garcia-Bellido:2017mdw}
\bibitem{Garcia-Bellido:2017mdw} 
  J.~Garc\'ia-Bellido and E.~Ruiz Morales,
  %``Primordial black holes from single field models of inflation,''
  arXiv:1702.03901 [astro-ph.CO].
  %%CITATION = ARXIV:1702.03901;%%
  %8 citations counted in INSPIRE as of 22 May 2017



%\cite{Motohashi:2017kbs}
\bibitem{Motohashi:2017kbs} 
  H.~Motohashi and W.~Hu,
  %``Primordial Black Holes and Slow-roll Violation,''
  arXiv:1706.06784 [astro-ph.CO].
  %%CITATION = ARXIV:1706.06784;%%


%\cite{Bezrukov:2017dyv}
\bibitem{Bezrukov:2017dyv} 
  F.~Bezrukov, M.~Pauly and J.~Rubio,
  %``On the robustness of the primordial power spectrum in renormalized Higgs inflation,''
  arXiv:1706.05007 [hep-ph].
  %%CITATION = ARXIV:1706.05007;%%
  %1 citations counted in INSPIRE as of 01 Jul 2017



%\cite{Deng:2016vzb}
\bibitem{Deng:2016vzb} 
  H.~Deng, J.~Garriga and A.~Vilenkin,
  %``Primordial black hole and wormhole formation by domain walls,''
  JCAP {\bf 1704}, no. 04, 050 (2017)
  %doi:10.1088/1475-7516/2017/04/050
  [arXiv:1612.03753 [gr-qc]].
  %%CITATION = doi:10.1088/1475-7516/2017/04/050;%%
  %2 citations counted in INSPIRE as of 22 May 2017
  

%\cite{Cotner:2016cvr}
\bibitem{Cotner:2016cvr} 
  E.~Cotner and A.~Kusenko,
  %``Primordial black holes from supersymmetry in the early universe,''
  arXiv:1612.02529 [astro-ph.CO].
  %%CITATION = ARXIV:1612.02529;%%
  %7 citations counted in INSPIRE as of 01 Jul 2017


%\cite{Cotner:2017tir}
\bibitem{Cotner:2017tir} 
  E.~Cotner and A.~Kusenko,
  %``Primordial black holes from scalar field evolution in the early universe,''
  arXiv:1706.09003 [astro-ph.CO].
  %%CITATION = ARXIV:1706.09003;%%


  
%\cite{Linde:2012bt}
\bibitem{Linde:2012bt} 
  A.~Linde, S.~Mooij and E.~Pajer,
  %``Gauge field production in supergravity inflation: Local non-Gaussianity and primordial black holes,''
  Phys.\ Rev.\ D {\bf 87}, no. 10, 103506 (2013)
  %doi:10.1103/PhysRevD.87.103506
  [arXiv:1212.1693 [hep-th]].
  %%CITATION = doi:10.1103/PhysRevD.87.103506;%%
  %39 citations counted in INSPIRE as of 23 Mar 2016


%\cite{Bugaev:2013fya}
\bibitem{Bugaev:2013fya} 
  E.~Bugaev and P.~Klimai,
  %``Axion inflation with gauge field production and primordial black holes,''
  Phys.\ Rev.\ D {\bf 90}, no. 10, 103501 (2014)
  %doi:10.1103/PhysRevD.90.103501
  [arXiv:1312.7435 [astro-ph.CO]].
  %%CITATION = doi:10.1103/PhysRevD.90.103501;%%
  %11 citations counted in INSPIRE as of 09 Jun 2016

%\cite{Erfani:2015rqv}
\bibitem{Erfani:2015rqv} 
  E.~Erfani,
  %``Primordial Black Holes Formation from Particle Production during Inflation,''
  JCAP {\bf 1604}, no. 04, 020 (2016)
  %doi:10.1088/1475-7516/2016/04/020
  [arXiv:1511.08470 [astro-ph.CO]].
  %%CITATION = doi:10.1088/1475-7516/2016/04/020;%%
  %2 citations counted in INSPIRE as of 19 May 2016

%\cite{Cheng:2016qzb}
\bibitem{Cheng:2016qzb} 
  S.~L.~Cheng, W.~Lee and K.~W.~Ng,
  %``Production of high stellar-mass primordial black holes in trapped inflation,''
  JHEP {\bf 1702}, 008 (2017)
  %doi:10.1007/JHEP02(2017)008
  [arXiv:1606.00206 [astro-ph.CO]].
  %%CITATION = doi:10.1007/JHEP02(2017)008;%%
  %8 citations counted in INSPIRE as of 29 Jun 2017

%\cite{Garcia-Bellido:2016dkw}
\bibitem{Garcia-Bellido:2016dkw} 
  J.~Garc\'ia-Bellido, M.~Peloso and C.~Unal,
  %``Gravitational waves at interferometer scales and primordial black holes in axion inflation,''
  JCAP {\bf 1612}, no. 12, 031 (2016)
  %doi:10.1088/1475-7516/2016/12/031
  [arXiv:1610.03763 [astro-ph.CO]].
  %%CITATION = doi:10.1088/1475-7516/2016/12/031;%%
  %12 citations counted in INSPIRE as of 25 May 2017


%\cite{Domcke:2017fix}
\bibitem{Domcke:2017fix} 
  V.~Domcke, F.~Muia, M.~Pieroni and L.~T.~Witkowski,
  %``PBH dark matter from axion inflation,''
  arXiv:1704.03464 [astro-ph.CO].
  %%CITATION = ARXIV:1704.03464;%%
  %2 citations counted in INSPIRE as of 22 May 2017


%\cite{Kannike:2017bxn}
\bibitem{Kannike:2017bxn} 
  K.~Kannike, L.~Marzola, M.~Raidal and H.~Veermäe,
  %``Single Field Double Inflation and Primordial Black Holes,''
  arXiv:1705.06225 [astro-ph.CO].
  %%CITATION = ARXIV:1705.06225;%%
  

%\cite{PinaAvelino:2005rm}
\bibitem{PinaAvelino:2005rm} 
  P.~Pina Avelino,
  %``Primordial black hole constraints on non-gaussian inflation models,''
  Phys.\ Rev.\ D {\bf 72}, 124004 (2005)
  %doi:10.1103/PhysRevD.72.124004
  [astro-ph/0510052].
  %%CITATION = doi:10.1103/PhysRevD.72.124004;%%
  %16 citations counted in INSPIRE as of 06 Jul 2017


 
    %\cite{Mollerach:2003nq}
\bibitem{Mollerach:2003nq} 
  S.~Mollerach, D.~Harari and S.~Matarrese,
  %``CMB polarization from secondary vector and tensor modes,''
  Phys.\ Rev.\ D {\bf 69}, 063002 (2004)
  %doi:10.1103/PhysRevD.69.063002
  [astro-ph/0310711].
  %%CITATION = doi:10.1103/PhysRevD.69.063002;%%
  %71 citations counted in INSPIRE as of 17 Jan 2017
  


  %\cite{Ananda:2006af}
\bibitem{Ananda:2006af} 
  K.~N.~Ananda, C.~Clarkson and D.~Wands,
  %``The Cosmological gravitational wave background from primordial density perturbations,''
  Phys.\ Rev.\ D {\bf 75}, 123518 (2007)
  %doi:10.1103/PhysRevD.75.123518
  [gr-qc/0612013].
  %%CITATION = doi:10.1103/PhysRevD.75.123518;%%
  %97 citations counted in INSPIRE as of 17 Jan 2017
  
 
  %\cite{Baumann:2007zm}
\bibitem{Baumann:2007zm} 
  D.~Baumann, P.~J.~Steinhardt, K.~Takahashi and K.~Ichiki,
  %``Gravitational Wave Spectrum Induced by Primordial Scalar Perturbations,''
  Phys.\ Rev.\ D {\bf 76}, 084019 (2007)
  %doi:10.1103/PhysRevD.76.084019
  [hep-th/0703290].
  %%CITATION = doi:10.1103/PhysRevD.76.084019;%%
  %106 citations counted in INSPIRE as of 17 Jan 2017




    
  
  %\cite{Clesse:2016ajp}
\bibitem{Clesse:2016ajp} 
  S.~Clesse and J.~Garc\'ia-Bellido,
  %``Detecting the gravitational wave background from primordial black hole dark matter,''
  arXiv:1610.08479 [astro-ph.CO].
  %%CITATION = ARXIV:1610.08479;%%
  %6 citations counted in INSPIRE as of 24 Dec 2016
  


  
\bibitem{Mandic:2016lcn} 
  V.~Mandic, S.~Bird and I.~Cholis,
  %``Stochastic Gravitational-Wave Background due to Primordial Binary Black Hole Mergers,''
  Phys.\ Rev.\ Lett.\  {\bf 117}, no. 20, 201102 (2016)
  %doi:10.1103/PhysRevLett.117.201102
  [arXiv:1608.06699 [astro-ph.CO]].
  %%CITATION = doi:10.1103/PhysRevLett.117.201102;%%
  


%\cite{Bartolo:2016ami}
\bibitem{Bartolo:2016ami} 
  N.~Bartolo {\it et al.},
  %``Science with the space-based interferometer LISA. IV: Probing inflation with gravitational waves,''
  JCAP {\bf 1612}, no. 12, 026 (2016)
  %doi:10.1088/1475-7516/2016/12/026
  [arXiv:1610.06481 [astro-ph.CO]].
  %%CITATION = doi:10.1088/1475-7516/2016/12/026;%%
  %22 citations counted in INSPIRE as of 30 Jun 2017



%\cite{Bird:2016dcv}
\bibitem{Bird:2016dcv} 
  S.~Bird, I.~Cholis, J.~B.~Muñoz, Y.~Ali-Haïmoud, M.~Kamionkowski, E.~D.~Kovetz, A.~Raccanelli and A.~G.~Riess,
  %``Did LIGO detect dark matter?,''
  Phys.\ Rev.\ Lett.\  {\bf 116}, no. 20, 201301 (2016)
% doi:10.1103/PhysRevLett.116.201301
  [arXiv:1603.00464 [astro-ph.CO]].
  %%CITATION = doi:10.1103/PhysRevLett.116.201301;%%
  %112 citations counted in INSPIRE as of 29 Jun 2017

%\cite{Sasaki:2016jop}
\bibitem{Sasaki:2016jop} 
  M.~Sasaki, T.~Suyama, T.~Tanaka and S.~Yokoyama,
  %``Primordial Black Hole Scenario for the Gravitational-Wave Event GW150914,''
  Phys.\ Rev.\ Lett.\  {\bf 117}, no. 6, 061101 (2016)
  %doi:10.1103/PhysRevLett.117.061101
  [arXiv:1603.08338 [astro-ph.CO]].
  %%CITATION = doi:10.1103/PhysRevLett.117.061101;%%
  %26 citations counted in INSPIRE as of 06 Oct 2016


 %\cite{Moore:2014lga}
\bibitem{Moore:2014lga} 
  C.~J.~Moore, R.~H.~Cole and C.~P.~L.~Berry,
  %``Gravitational-wave sensitivity curves,''
  Class.\ Quant.\ Grav.\  {\bf 32}, no. 1, 015014 (2015)
  %doi:10.1088/0264-9381/32/1/015014
  [arXiv:1408.0740 [gr-qc]].
  %%CITATION = doi:10.1088/0264-9381/32/1/015014;%%
  %46 citations counted in INSPIRE as of 05 Dec 2016



  
%\cite{Inomata:2016rbd}
\bibitem{Inomata:2016rbd} 
  K.~Inomata, M.~Kawasaki, K.~Mukaida, Y.~Tada and T.~T.~Yanagida,
  %``Inflationary primordial black holes for the LIGO gravitational wave events and pulsar timing array experiments,''
  arXiv:1611.06130 [astro-ph.CO].
  %%CITATION = ARXIV:1611.06130;%%
  %10 citations counted in INSPIRE as of 02 Jun 2017

%\cite{Orlofsky:2016vbd}
\bibitem{Orlofsky:2016vbd} 
  N.~Orlofsky, A.~Pierce and J.~D.~Wells,
  %``Inflationary theory and pulsar timing investigations of primordial black holes and gravitational waves,''
  Phys.\ Rev.\ D {\bf 95}, no. 6, 063518 (2017)
  %doi:10.1103/PhysRevD.95.063518
  [arXiv:1612.05279 [astro-ph.CO]].
  %%CITATION = doi:10.1103/PhysRevD.95.063518;%%
  %6 citations counted in INSPIRE as of 26 Jun 2017


 
 %\cite{Nakama:2016gzw}
\bibitem{Nakama:2016gzw} 
  T.~Nakama, J.~Silk and M.~Kamionkowski,
  %``Stochastic gravitational waves associated with the formation of primordial black holes,''
  Phys.\ Rev.\ D {\bf 95}, no. 4, 043511 (2017)
  %doi:10.1103/PhysRevD.95.043511
  [arXiv:1612.06264 [astro-ph.CO]].
  %%CITATION = doi:10.1103/PhysRevD.95.043511;%%
  %5 citations counted in INSPIRE as of 26 Jun 2017



%\cite{Garcia-Bellido:2017qal}
\bibitem{Garcia-Bellido:2017qal} 
  J.~Garc\'ia-Bellido and S.~Nesseris,
%  ``Gravitational wave bursts from Primordial Black Hole hyperbolic encounters,''
  arXiv:1706.02111 [astro-ph.CO].
  %%CITATION = ARXIV:1706.02111;%%


%\cite{Nakama:2016kfq}
\bibitem{Nakama:2016kfq} 
  T.~Nakama, T.~Suyama and J.~Yokoyama,
  %``Supermassive black holes formed by direct collapse of inflationary perturbations,''
  Phys.\ Rev.\ D {\bf 94}, no. 10, 103522 (2016)
  doi:10.1103/PhysRevD.94.103522
  [arXiv:1609.02245 [gr-qc]].
  %%CITATION = doi:10.1103/PhysRevD.94.103522;%%
  %3 citations counted in INSPIRE as of 06 Jul 2017


%\cite{Kogut:2011xw}
\bibitem{PIXIE} 
  A.~Kogut {\it et al.},
  %``The Primordial Inflation Explorer (PIXIE): A Nulling Polarimeter for Cosmic Microwave Background Observations,''
  JCAP {\bf 1107}, 025 (2011)
  %doi:10.1088/1475-7516/2011/07/025
  [arXiv:1105.2044 [astro-ph.CO]].
  %%CITATION = doi:10.1088/1475-7516/2011/07/025;%%
  %272 citations counted in INSPIRE as of 05 Jul 2017


%\cite{Andre:2013afa}
\bibitem{PRISM} 
  P.~Andre {\it et al.} [PRISM Collaboration],
  %``PRISM (Polarized Radiation Imaging and Spectroscopy Mission): A White Paper on the Ultimate Polarimetric Spectro-Imaging of the Microwave and Far-Infrared Sky,''
  arXiv:1306.2259 [astro-ph.CO].
  %%CITATION = ARXIV:1306.2259;%%
  %105 citations counted in INSPIRE as of 05 Jul 2017





%\cite{Barnacka:2012bm}
\bibitem{Barnacka:2012bm} 
  A.~Barnacka, J.~F.~Glicenstein and R.~Moderski,
  %``New constraints on primordial black holes abundance from femtolensing of gamma-ray bursts,''
  Phys.\ Rev.\ D {\bf 86}, 043001 (2012)
  %doi:10.1103/PhysRevD.86.043001
  [arXiv:1204.2056 [astro-ph.CO]].
  %%CITATION = doi:10.1103/PhysRevD.86.043001;%%
  %29 citations counted in INSPIRE as of 14 Sep 2016




%\cite{Capela:2014ita}
\bibitem{Capela:2014ita} 
  F.~Capela, M.~Pshirkov and P.~Tinyakov,
  %``Adiabatic contraction revisited: implications for primordial black holes,''
  Phys.\ Rev.\ D {\bf 90}, no. 8, 083507 (2014)
  %doi:10.1103/PhysRevD.90.083507
  [arXiv:1403.7098 [astro-ph.CO]].
  %%CITATION = doi:10.1103/PhysRevD.90.083507;%%
  %13 citations counted in INSPIRE as of 14 Sep 2016


%\cite{Kawasaki:2016pql}
\bibitem{Kawasaki:2016pql} 
  M.~Kawasaki, A.~Kusenko, Y.~Tada and T.~T.~Yanagida,
  %``PBH Dark Matter in Supergravity Inflation Models,''
  arXiv:1606.07631 [astro-ph.CO].
  %%CITATION = ARXIV:1606.07631;%%
  %1 citations counted in INSPIRE as of 07 Oct 2016








%\cite{Griest:2013aaa}
\bibitem{Griest:2013aaa} 
  K.~Griest, A.~M.~Cieplak and M.~J.~Lehner,
  %``Experimental Limits on Primordial Black Hole Dark Matter from the First 2 yr of Kepler Data,''
  Astrophys.\ J.\  {\bf 786}, no. 2, 158 (2014)
  %doi:10.1088/0004-637X/786/2/158
  [arXiv:1307.5798 [astro-ph.CO]].
  %%CITATION = doi:10.1088/0004-637X/786/2/158;%%
  %24 citations counted in INSPIRE as of 14 Sep 2016



%\cite{Tisserand:2006zx}
\bibitem{Tisserand:2006zx} 
  P.~Tisserand {\it et al.} [EROS-2 Collaboration],
  %``Limits on the Macho Content of the Galactic Halo from the EROS-2 Survey of the Magellanic Clouds,''
  Astron.\ Astrophys.\  {\bf 469}, 387 (2007)
  %doi:10.1051/0004-6361:20066017
  [astro-ph/0607207].
  %%CITATION = doi:10.1051/0004-6361:20066017;%%
  %213 citations counted in INSPIRE as of 14 Sep 2016

%\cite{Alcock:1998fx}
\bibitem{Alcock:1998fx} 
  C.~Alcock {\it et al.} [MACHO and EROS Collaborations],
  %``EROS and MACHO combined limits on planetary mass dark matter in the galactic halo,''
  Astrophys.\ J.\  {\bf 499}, L9 (1998)
  %doi:10.1086/311355
  [astro-ph/9803082].
  %%CITATION = doi:10.1086/311355;%%
  %103 citations counted in INSPIRE as of 14 Sep 2016




\bibitem{Brandt:2016aco} 
  T.~D.~Brandt,
  %``Constraints on MACHO Dark Matter from Compact Stellar Systems in Ultra-Faint Dwarf Galaxies,''
  Astrophys.\ J.\  {\bf 824}, no. 2, L31 (2016)
  %doi:10.3847/2041-8205/824/2/L31
  [arXiv:1605.03665 [astro-ph.GA]].
  %%CITATION = doi:10.3847/2041-8205/824/2/L31;%%
  %3 citations counted in INSPIRE as of 24 Sep 2016


\bibitem{Li:2016utv} 
  T.~S.~Li {\it et al.} [DES Collaboration],
  %``Farthest Neighbor: The Distant Milky Way Satellite Eridanus II,''
  Astrophys.\ J.\  {\bf 838}, no. 1, 8 (2017)
  %doi:10.3847/1538-4357/aa6113
  [arXiv:1611.05052 [astro-ph.GA]].
  %%CITATION = doi:10.3847/1538-4357/aa6113;%%
  %7 citations counted in INSPIRE as of 09 Jun 2017
  
  
  %\cite{Quinn:2009zg}
\bibitem{Quinn:2009zg} 
  D.~P.~Quinn, M.~I.~Wilkinson, M.~J.~Irwin, J.~Marshall, A.~Koch and V.~Belokurov,
  %``On the Reported Death of the MACHO Era,''
  Mon.\ Not.\ Roy.\ Astron.\ Soc.\  {\bf 396}, 11 (2009)
  %doi:10.1111/j.1745-3933.2009.00652.x
  [arXiv:0903.1644 [astro-ph.GA]].
  %%CITATION = doi:10.1111/j.1745-3933.2009.00652.x;%%
  %17 citations counted in INSPIRE as of 14 Sep 2016
  
  



\bibitem{Ricotti:2007au} 
  M.~Ricotti, J.~P.~Ostriker and K.~J.~Mack,
  %``Effect of Primordial Black Holes on the Cosmic Microwave Background and Cosmological Parameter Estimates,''
  Astrophys.\ J.\  {\bf 680}, 829 (2008)
  %doi:10.1086/587831
  [arXiv:0709.0524 [astro-ph]].
  %%CITATION = doi:10.1086/587831;%%
  %81 citations counted in INSPIRE as of 24 Sep 2016




  
  %\cite{Ali-Haimoud:2016mbv}
\bibitem{Ali-Haimoud:2016mbv} 
  Y.~Ali-Haïmoud and M.~Kamionkowski,
  %``Cosmic microwave background limits on accreting primordial black holes,''
  arXiv:1612.05644 [astro-ph.CO].
  %%CITATION = ARXIV:1612.05644;%%
  %3 citations counted in INSPIRE as of 27 Dec 2016
  

  %\cite{Lyth:2012yp}
\bibitem{Lyth:2012yp} 
  D.~H.~Lyth,
  %``The hybrid inflation waterfall and the primordial curvature perturbation,''
  JCAP {\bf 1205}, 022 (2012)
  %doi:10.1088/1475-7516/2012/05/022
  [arXiv:1201.4312 [astro-ph.CO]].
  %%CITATION = doi:10.1088/1475-7516/2012/05/022;%%
  %43 citations counted in INSPIRE as of 15 Sep 2016
  
  
  %\cite{Byrnes:2012yx}
\bibitem{Byrnes:2012yx} 
  C.~T.~Byrnes, E.~J.~Copeland and A.~M.~Green,
  %``Primordial black holes as a tool for constraining non-Gaussianity,''
  Phys.\ Rev.\ D {\bf 86}, 043512 (2012)
  %doi:10.1103/PhysRevD.86.043512
  [arXiv:1206.4188 [astro-ph.CO]].
  %%CITATION = doi:10.1103/PhysRevD.86.043512;%%
  %28 citations counted in INSPIRE as of 15 Sep 2016



%\cite{Namba:2015gja}
\bibitem{Namba:2015gja} 
  R.~Namba, M.~Peloso, M.~Shiraishi, L.~Sorbo and C.~Unal,
  %``Scale-dependent gravitational waves from a rolling axion,''
  JCAP {\bf 1601}, no. 01, 041 (2016)
  %doi:10.1088/1475-7516/2016/01/041
  [arXiv:1509.07521 [astro-ph.CO]].
  %%CITATION = doi:10.1088/1475-7516/2016/01/041;%%
  %6 citations counted in INSPIRE as of 23 Mar 2016







  
%\cite{Ade:2015lrj}
\bibitem{Ade:2015lrj} 
  P.~A.~R.~Ade {\it et al.}  [Planck Collaboration],
  %``Planck 2015 results. XX. Constraints on inflation,''
  arXiv:1502.02114 [astro-ph.CO].
  %%CITATION = ARXIV:1502.02114;%%
  %53 citations counted in INSPIRE as of 28 Mar 2015





%\cite{Arzoumanian:2015liz}
\bibitem{Arzoumanian:2015liz} 
  Z.~Arzoumanian {\it et al.} [NANOGrav Collaboration],
  %``The NANOGrav Nine-year Data Set: Limits on the Isotropic Stochastic Gravitational Wave Background,''
  Astrophys.\ J.\  {\bf 821}, no. 1, 13 (2016)
  %doi:10.3847/0004-637X/821/1/13
  [arXiv:1508.03024 [astro-ph.GA]].
  %%CITATION = doi:10.3847/0004-637X/821/1/13;%%
  %52 citations counted in INSPIRE as of 05 Dec 2016


%\cite{Lentati:2015qwp}
\bibitem{Lentati:2015qwp} 
  L.~Lentati {\it et al.},
  %``European Pulsar Timing Array Limits On An Isotropic Stochastic Gravitational-Wave Background,''
  Mon.\ Not.\ Roy.\ Astron.\ Soc.\  {\bf 453}, no. 3, 2576 (2015)
  %doi:10.1093/mnras/stv1538
  [arXiv:1504.03692 [astro-ph.CO]].
  %%CITATION = doi:10.1093/mnras/stv1538;%%
  %67 citations counted in INSPIRE as of 05 Dec 2016


%\cite{Shannon:2015ect}
\bibitem{Shannon:2015ect} 
  R.~M.~Shannon {\it et al.},
  %``Gravitational waves from binary supermassive black holes missing in pulsar observations,''
  Science {\bf 349}, no. 6255, 1522 (2015)
  %doi:10.1126/science.aab1910
  [arXiv:1509.07320 [astro-ph.CO]].
  %%CITATION = doi:10.1126/science.aab1910;%%
  %61 citations counted in INSPIRE as of 05 Dec 2016
 
 

%\cite{Zhao:2013bba}
\bibitem{Zhao:2013bba} 
  W.~Zhao, Y.~Zhang, X.~P.~You and Z.~H.~Zhu,
  %``Constraints of relic gravitational waves by pulsar timing arrays: Forecasts for the FAST and SKA projects,''
  Phys.\ Rev.\ D {\bf 87}, no. 12, 124012 (2013)
  %doi:10.1103/PhysRevD.87.124012
  [arXiv:1303.6718 [astro-ph.CO]].
  %%CITATION = doi:10.1103/PhysRevD.87.124012;%%
  %19 citations counted in INSPIRE as of 06 Dec 2016
  


%\cite{Chluba:2015bqa}
\bibitem{Chluba:2015bqa} 
  J.~Chluba, J.~Hamann and S.~P.~Patil,
  %``Features and New Physical Scales in Primordial Observables: Theory and Observation,''
  Int.\ J.\ Mod.\ Phys.\ D {\bf 24}, no. 10, 1530023 (2015)
  doi:10.1142/S0218271815300232
  [arXiv:1505.01834 [astro-ph.CO]].
  %%CITATION = doi:10.1142/S0218271815300232;%%
  %46 citations counted in INSPIRE as of 04 Jul 2017


%\cite{Nakama:2017ohe}
\bibitem{Nakama:2017ohe} 
  T.~Nakama, J.~Chluba and M.~Kamionkowski,
  %``Shedding light on the small-scale crisis with CMB spectral distortions,''
  Phys.\ Rev.\ D {\bf 95}, 121302
  [Phys.\ Rev.\ D {\bf 95}, 121302 (2017)]
  doi:10.1103/PhysRevD.95.121302
  [arXiv:1703.10559 [astro-ph.CO]].
  %%CITATION = doi:10.1103/PhysRevD.95.121302;%%
  %2 citations counted in INSPIRE as of 04 Jul 2017


%\cite{Mather:1993ij}
\bibitem{Mather:1993ij} 
  J.~C.~Mather {\it et al.},
  %``Measurement of the Cosmic Microwave Background spectrum by the COBE FIRAS instrument,''
  Astrophys.\ J.\  {\bf 420}, 439 (1994).
  doi:10.1086/173574
  %%CITATION = doi:10.1086/173574;%%
  %457 citations counted in INSPIRE as of 04 Jul 2017


%\cite{Fixsen:1996nj}
\bibitem{Fixsen:1996nj} 
  D.~J.~Fixsen, E.~S.~Cheng, J.~M.~Gales, J.~C.~Mather, R.~A.~Shafer and E.~L.~Wright,
  %``The Cosmic Microwave Background spectrum from the full COBE FIRAS data set,''
  Astrophys.\ J.\  {\bf 473}, 576 (1996)
  doi:10.1086/178173
  [astro-ph/9605054].
  %%CITATION = doi:10.1086/178173;%%
  %818 citations counted in INSPIRE as of 04 Jul 2017

%\cite{Abitbol:2017vwa}
\bibitem{Abitbol:2017vwa} 
  M.~H.~Abitbol, J.~Chluba, J.~C.~Hill and B.~R.~Johnson,
  %``Prospects for Measuring Cosmic Microwave Background Spectral Distortions in the Presence of Foregrounds,''
  doi:10.1093/mnras/stx1653
  arXiv:1705.01534 [astro-ph.CO].
  %%CITATION = doi:10.1093/mnras/stx1653;%%
  %2 citations counted in INSPIRE as of 04 Jul 2017




%\cite{Carr:1975qj}
\bibitem{Carr:1975qj} 
  B.~J.~Carr,
  %``The Primordial black hole mass spectrum,''
  Astrophys.\ J.\  {\bf 201}, 1 (1975).
  %doi:10.1086/153853
  %%CITATION = doi:10.1086/153853;%%
  %377 citations counted in INSPIRE as of 15 Sep 2016


 


%\cite{Nakama:2013ica}
\bibitem{Nakama:2013ica} 
  T.~Nakama, T.~Harada, A.~G.~Polnarev and J.~Yokoyama,
  %``Identifying the most crucial parameters of the initial curvature profile for primordial black hole formation,''
  JCAP {\bf 1401}, 037 (2014)
  %doi:10.1088/1475-7516/2014/01/037
  [arXiv:1310.3007 [gr-qc], arXiv:1310.3007].
  %%CITATION = doi:10.1088/1475-7516/2014/01/037;%%
  %20 citations counted in INSPIRE as of 27 Dec 2016
  
  

 
  
  %\cite{Shibata:1999zs}
\bibitem{Shibata:1999zs} 
  M.~Shibata and M.~Sasaki,
  %``Black hole formation in the Friedmann universe: Formulation and computation in numerical relativity,''
  Phys.\ Rev.\ D {\bf 60}, 084002 (1999)
  %doi:10.1103/PhysRevD.60.084002
  [gr-qc/9905064].
  %%CITATION = doi:10.1103/PhysRevD.60.084002;%%
  %105 citations counted in INSPIRE as of 27 Dec 2016
  
  
  %\cite{Polnarev:2006aa}
\bibitem{Polnarev:2006aa} 
  A.~G.~Polnarev and I.~Musco,
  %``Curvature profiles as initial conditions for primordial black hole formation,''
  Class.\ Quant.\ Grav.\  {\bf 24}, 1405 (2007)
  %doi:10.1088/0264-9381/24/6/003
  [gr-qc/0605122].
  %%CITATION = doi:10.1088/0264-9381/24/6/003;%%
  %32 citations counted in INSPIRE as of 27 Dec 2016
  
 
    
%\cite{Khlopov:2008qy}
\bibitem{Khlopov:2008qy} 
  M.~Y.~Khlopov,
  %``Primordial Black Holes,''
  Res.\ Astron.\ Astrophys.\  {\bf 10}, 495 (2010)
  %doi:10.1088/1674-4527/10/6/001
  [arXiv:0801.0116 [astro-ph]].
  %%CITATION = doi:10.1088/1674-4527/10/6/001;%%
  %91 citations counted in INSPIRE as of 20 Dec 2016


    
 %\cite{Carr:2016drx}
\bibitem{Carr:2016drx} 
  B.~Carr, F.~Kuhnel and M.~Sandstad,
  %``Primordial Black Holes as Dark Matter,''
  Phys.\ Rev.\ D {\bf 94}, no. 8, 083504 (2016)
  %doi:10.1103/PhysRevD.94.083504
  [arXiv:1607.06077 [astro-ph.CO]].
  %%CITATION = doi:10.1103/PhysRevD.94.083504;%%
  %83 citations counted in INSPIRE as of 05 Jul 2017

    
%\cite{Inomata:2017okj}
\bibitem{Inomata:2017okj} 
  K.~Inomata, M.~Kawasaki, K.~Mukaida, Y.~Tada and T.~T.~Yanagida,
  %``Inflationary Primordial Black Holes as All Dark Matter,''
  arXiv:1701.02544 [astro-ph.CO].
  %%CITATION = ARXIV:1701.02544;%%
  %8 citations counted in INSPIRE as of 29 Jun 2017

  
%\cite{Carr:2009jm}
\bibitem{Carr:2009jm} 
  B.~J.~Carr, K.~Kohri, Y.~Sendouda and J.~Yokoyama,
  %``New cosmological constraints on primordial black holes,''
  Phys.\ Rev.\ D {\bf 81}, 104019 (2010)
  %doi:10.1103/PhysRevD.81.104019
  [arXiv:0912.5297 [astro-ph.CO]].
  %%CITATION = doi:10.1103/PhysRevD.81.104019;%%
  %240 citations counted in INSPIRE as of 29 Aug 2016


%\cite{Cuesta:2015mqa}
\bibitem{Cuesta:2015mqa} 
  A.~J.~Cuesta {\it et al.},
  %``The clustering of galaxies in the SDSS-III Baryon Oscillation Spectroscopic Survey: Baryon Acoustic Oscillations in the correlation function of LOWZ and CMASS galaxies in Data Release 12,''
  Mon.\ Not.\ Roy.\ Astron.\ Soc.\  {\bf 457}, no. 2, 1770 (2016)
  %doi:10.1093/mnras/stw066
  [arXiv:1509.06371 [astro-ph.CO]].
  %%CITATION = doi:10.1093/mnras/stw066;%%
  %40 citations counted in INSPIRE as of 06 Oct 2016




%\cite{Bringmann:2011ut}
\bibitem{Bringmann:2011ut} 
  T.~Bringmann, P.~Scott and Y.~Akrami,
  %``Improved constraints on the primordial power spectrum at small scales from ultracompact minihalos,''
  Phys.\ Rev.\ D {\bf 85}, 125027 (2012),
  %doi:10.1103/PhysRevD.85.125027
  [arXiv:1110.2484 [astro-ph.CO]].
  %%CITATION = doi:10.1103/PhysRevD.85.125027;%%
  %69 citations counted in INSPIRE as of 10 Feb 2017



%\cite{Emami:2017fiy}
\bibitem{Emami:2017fiy} 
  R.~Emami and G.~Smoot,
  %``Observational Constraints on the Primordial Curvature Power Spectrum,''
  arXiv:1705.09924 [astro-ph.CO].
  %%CITATION = ARXIV:1705.09924;%%
  %1 citations counted in INSPIRE as of 29 Jun 2017

  

%\cite{Ade:2015xua}
\bibitem{Ade:2015xua} 
  P.~A.~R.~Ade {\it et al.} [Planck Collaboration],
  %``Planck 2015 results. XIII. Cosmological parameters,''
  Astron.\ Astrophys.\  {\bf 594}, A13 (2016)
  %doi:10.1051/0004-6361/201525830
  [arXiv:1502.01589 [astro-ph.CO]].
  %%CITATION = doi:10.1051/0004-6361/201525830;%%
  %2208 citations counted in INSPIRE as of 04 Oct 2016



%\cite{Peloso:2016gqs}
\bibitem{Peloso:2016gqs} 
  M.~Peloso, L.~Sorbo and C.~Unal,
  %``Rolling axions during inflation: perturbativity and signatures,''
  JCAP {\bf 1609}, no. 09, 001 (2016)
  %doi:10.1088/1475-7516/2016/09/001
  [arXiv:1606.00459 [astro-ph.CO]].
  %%CITATION = doi:10.1088/1475-7516/2016/09/001;%%
  %4 citations counted in INSPIRE as of 12 Sep 2016







\end{thebibliography}
\end{document}